\documentclass[conference]{IEEEtran}
\IEEEoverridecommandlockouts

\usepackage[authormarkup=none]{changes}
\usepackage{listings}
\usepackage{tikz}
\usepackage{verbatimbox}
\usepackage{multirow}
\usepackage{array}
\usepackage[utf8]{inputenc}
\usepackage{booktabs}
\usepackage{makecell}
\usepackage{enumitem}
\usepackage{float}
\usepackage{amsthm}
\usepackage{subfigure}
\usepackage{algorithm}
\usepackage{algpseudocode}
\usepackage{cite}
\usepackage{amsmath,amssymb,amsfonts}
\usepackage{graphicx}
\usepackage{textcomp}
\usepackage{xcolor}
\usepackage{balance}

\graphicspath{{figures/}}

\newtheorem{definition}{Definition}

\newcommand{\approachname}{\texttt{GraphQE}}
\newcommand{\dataset}{\texttt{CyEqSet}}
\newcommand{\uexpg}{\emph{G-expression}}
\newcommand{\uexpt}{\emph{U-expression}}

\newcommand{\figref}[1]{Fig.~\ref{#1}}
\newcommand{\tabref}[1]{Table~\ref{#1}}
\newcommand{\secref}[1]{Section~\ref{#1}}

\newcommand{\listref}[1]{Listing~\ref{#1}}

\newcommand{\review}[1]{} % Hide all review comments

\newcommand{\revision}[1]{#1}
\newcommand{\id}[1]{} % Hide all question IDs

\newcounter{finding}

\newcommand{\finding}[2]{\refstepcounter{finding} \label{finding:#1}
  \begin{center}
  \begin{tikzpicture}%
    \node[rectangle, draw=black, top color=black!3, bottom
    color=black!3, rounded corners=2pt, inner xsep=5pt, inner
    ysep=6pt, outer ysep=10pt]{
    \begin{minipage}{0.95\columnwidth}
      \textit{#2}
    \end{minipage}};%
  \end{tikzpicture}%
  \end{center}
}

\definechangesauthor[color=cyan]{tl}
\newcommand{\tl}[1]{\textcolor{cyan}{[tl: #1]}}
\newcommand{\tladd}[1]{\added[id=tl]{#1}}
\newcommand{\tldelete}[1]{\deleted[id=tl]{#1}}
\newcommand{\tlreplace}[2]{\replaced[id=tl]{#1}{#2}}

\definechangesauthor[color=orange]{yingying}
\newcommand{\yingying}[1]{\textcolor{orange}{[yingying: #1]}}

\definechangesauthor[color=red]{wensheng}
\newcommand{\wensheng}[1]{\textcolor{red}{[wensheng: #1]}}
\newcommand{\wsadd}[1]{\added[id=wensheng]{#1}}

\definechangesauthor[color=cyan]{xudong}

\definechangesauthor[color=magenta]{jiansen}

\definechangesauthor[color=brown]{ziyu}

\newcommand*\circled[1]{\tikz[baseline=-0.5ex]{
  \node[shape=circle,draw,inner sep=4pt] (char) {};}}

\begin{document}

\title{Proving Cypher Query Equivalence
\thanks{Wensheng Dou is the corresponding author.}}

\author{\IEEEauthorblockN{Lei Tang, Wensheng Dou, Yingying Zheng, Lijie Xu, Wei
Wang, Jun Wei, Tao Huang} 
\IEEEauthorblockA{Institute of Software Chinese Academy of Sciences, China \\
\{tanglei20, wsdou, zhengyingying14, xulijie09, wangwei, wj, tao\}@otcaix.iscas.ac.cn}}

\author{
  \IEEEauthorblockN{
    Lei Tang\IEEEauthorrefmark{1}\IEEEauthorrefmark{2},
    Wensheng Dou\IEEEauthorrefmark{1}\IEEEauthorrefmark{2}\IEEEauthorrefmark{3}\IEEEauthorrefmark{4},
    Yingying Zheng\IEEEauthorrefmark{1}\IEEEauthorrefmark{2},
    Lijie Xu\IEEEauthorrefmark{1}\IEEEauthorrefmark{2},
    Wei Wang\IEEEauthorrefmark{1}\IEEEauthorrefmark{2}\IEEEauthorrefmark{3}\IEEEauthorrefmark{4},
    Jun Wei\IEEEauthorrefmark{1}\IEEEauthorrefmark{2}\IEEEauthorrefmark{3}\IEEEauthorrefmark{4},
    Tao Huang\IEEEauthorrefmark{1}\IEEEauthorrefmark{2}
  }
  \IEEEauthorblockA{\IEEEauthorrefmark{1}\textit{Key Lab of System Software, State Key Laboratory of Computer Science, Institute of Software,
  Chinese Academy of Sciences}}
  \IEEEauthorblockA{\IEEEauthorrefmark{2}\textit{University of Chinese Academy
  of Sciences, Beijing}}
  \IEEEauthorblockA{\IEEEauthorrefmark{3}\textit{University of Chinese Academy
  of Sciences, Nanjing}, \IEEEauthorrefmark{4}\textit{Nanjing Institute of
  Software Technology} \\
  \{tanglei20, wsdou, zhengyingying14, xulijie09, wangwei, wj, tao\}@otcaix.iscas.ac.cn}
}

\maketitle

\definecolor{keywordcolor}{rgb}{0.13,0.29,0.53}
\definecolor{stringcolor}{rgb}{0.31,0.60,0.02}
\definecolor{deepblue}{RGB}{0, 0, 139}
\definecolor{codegreen}{rgb}{0,0.6,0}
\definecolor{codegray}{rgb}{0.5,0.5,0.5}
\definecolor{codepurple}{rgb}{0.58,0,0.82}

\lstdefinestyle{cypherstyle}{ 
  numberstyle=\footnotesize\color{codegray},
  stringstyle=\color{codepurple},
  commentstyle=\color{gray},         
  keywordstyle=[1]\color{blue}, 
  % keywordstyle=[2]\color{ACMPurple}\bfseries,
  basicstyle=\ttfamily\selectfont\footnotesize,
  breakatwhitespace=false, 
  breaklines=true,                 
  captionpos=b,                    
  keepspaces=true,                 
  numbers=left,                    
  numbersep=5pt,                  
  showspaces=false,                
  showstringspaces=false,
  showtabs=false,                  
  tabsize=2,
  frame=single,                    
  framesep=2pt,                    
  framerule=0.5pt,                 
  xleftmargin=10pt,                
  xrightmargin=5pt,
  captionpos=b,
  morecomment=[s]{/*}{*/},
  keywords=[1]{MATCH, RETURN, OPTIONAL,
  WHERE, WITH, CALL,SELECT,VALUES,FROM, UNION,JOIN, ALL, UNWIND, EXISTS,ON,
  PREFIX, has,hasLabel,out,values},
  % keywords=[2]{SELECT, FROM, RETURN,
  % WHERE, VALUES, UNION, ALL, JOIN,ON}
}

\lstdefinestyle{sqlstyle}{
    language=SQL,         
    numberstyle=\tiny\color{codegray},  
    commentstyle=\color{gray}, 
    keywordstyle=\color{purple}\bfseries,
    stringstyle=\color{green},
    basicstyle=\ttfamily\footnotesize,
    breakatwhitespace=false, 
    breaklines=true,                 
    captionpos=b,                    
    keepspaces=true,                 
    numbers=left,                    
    numbersep=10pt,                  
    showspaces=false,                
    showstringspaces=false,
    showtabs=false,                  
    tabsize=2,
    frame=single,                    
    framesep=2pt,                    
    framerule=0.5pt,                 
    xleftmargin=10pt,                
    xrightmargin=5pt,
    captionpos=b,
    morecomment=[s]{/*}{*/}
}

\newcommand{\join}{%
  \mathbin{%
    \begin{tikzpicture}[baseline=-0.5ex]
      \draw[line width=0.5pt] (0,0) -- (0.1,0.1);
      \draw[line width=0.5pt] (0,0) -- (0.1,-0.1);
      \draw[line width=0.5pt] (0.1,0.1) -- (0.1,-0.1);
      \draw[line width=0.5pt] (-0.1,-0.1) -- (0,0);
      \draw[line width=0.5pt] (-0.1,0.1) -- (0,0);
      \draw[line width=0.5pt] (-0.1,-0.1) -- (-0.1,0.1);
    \end{tikzpicture}%
  }
}

\newcommand{\outerjoin}{%
  \mathbin{%
    \begin{tikzpicture}[baseline=-0.5ex]
      \draw[line width=0.5pt] (0,0) -- (0.1,0.1);
      \draw[line width=0.5pt] (0,0) -- (0.1,-0.1);
      \draw[line width=0.5pt] (0.1,0.1) -- (0.1,-0.1);
      \draw[line width=0.5pt] (-0.1,-0.1) -- (0,0);
      \draw[line width=0.5pt] (-0.1,0.1) -- (0,0);
      \draw[line width=0.5pt] (-0.1,-0.1) -- (-0.1,0.1);
      \draw[line width=0.5pt] (-0.2,0.1) -- (-0.1,0.1);
      \draw[line width=0.5pt] (-0.2,-0.1) -- (-0.1,-0.1);
    \end{tikzpicture}%
  }
}
% %
% % The abstract is a short summary of the work to be presented in the
% % article.
\begin{abstract}

Graph database systems store graph data as nodes and relationships, and
utilize graph query languages (e.g., Cypher) for efficiently querying graph
data. Proving the equivalence of graph queries is an important foundation for
optimizing graph query performance, ensuring graph query reliability, etc. Although
researchers have proposed many SQL query equivalence provers for relational
database systems, these provers cannot be directly applied to prove the
equivalence of graph queries. The difficulty lies in the fact that graph query
languages (e.g., Cypher) adopt significantly different data models (property
graph model vs. relational model) and query patterns (graph pattern matching vs.
tabular tuple calculus) from SQL.

In this paper, we propose \approachname{}, an automated prover to determine
whether two Cypher queries are semantically equivalent.
We design a U-semiring based Cypher algebraic representation to model the
semantics of Cypher queries. Our Cypher algebraic representation is built on the
algebraic structure of unbounded semirings, and can sufficiently express nodes
and relationships in property graphs and complex Cypher queries. Then,
determining the equivalence of two Cypher queries is transformed into
determining the equivalence of the corresponding Cypher algebraic representations, which can be verified by SMT
solvers. To evaluate the effectiveness of \approachname{}, we construct a
dataset consisting of 148 pairs of equivalent Cypher queries. Among them, we
have successfully proven 138 pairs of equivalent Cypher queries, demonstrating the effectiveness of \approachname{}. 

\end{abstract}

\maketitle

\iffalse
\fi

% \pagestyle{\vldbpagestyle}
% \begingroup\small\noindent\raggedright\textbf{PVLDB Reference Format:}\\
% \vldbauthors. \vldbtitle. PVLDB, \vldbvolume(\vldbissue): \vldbpages, \vldbyear.\\
% \href{https://doi.org/\vldbdoi}{doi:\vldbdoi}
% \endgroup
% \begingroup
% \renewcommand\thefootnote{}\footnote{\noindent
% This work is licensed under the Creative Commons BY-NC-ND 4.0 International License. Visit \url{https://creativecommons.org/licenses/by-nc-nd/4.0/} to view a copy of this license. For any use beyond those covered by this license, obtain permission by emailing \href{mailto:info@vldb.org}{info@vldb.org}. Copyright is held by the owner/author(s). Publication rights licensed to the VLDB Endowment. \\
% \raggedright Proceedings of the VLDB Endowment, Vol. \vldbvolume, No. \vldbissue\ %
% ISSN 2150-8097. \\
% \href{https://doi.org/\vldbdoi}{doi:\vldbdoi} \\
% }\addtocounter{footnote}{-1}\endgroup
% %%% VLDB block end %%%

% %%% do not modify the following VLDB block %%
% %%% VLDB block start %%%
% \ifdefempty{\vldbavailabilityurl}{}{
% \vspace{.3cm}
% \begingroup\small\noindent\raggedright\textbf{PVLDB Artifact Availability:}\\
% The source code, data, and/or other artifacts have been made available at
% \url{https://github.com/choeoe/CyEq}.
% \endgroup
% }
\section{Introduction}

Graph database systems (GDBs) are designed to efficiently store and retrieve
graph data. Graph storage technologies \cite{mpinda2015evaluation,
de2017smart} have developed rapidly, and many GDBs have emerged, e.g., Neo4j
\cite{kemper2015beginning}, Microsoft Azure Cosmos \cite{azurecosmos}, ArangoDB
\cite{arangodb} and Memgraph \cite{memgraph2023}. GDBs have been widely used
in many applications, e.g., knowledge graphs \cite{wang2020covid}, fraud
detection \cite{ren21frauddetection}, molecular and cell biology
\cite{eckman2006graph}, and social networks \cite{fan2019social}. Recent reports
\cite{dbengines2023graph, imarc2024graph} also show that GDBs have gained
almost 1,000\% popularity growth since 2013 and reached a market of around 1.7
billion dollars by 2023.

\begin{figure}[t]
  \centering
  \includegraphics[width=0.9\linewidth]{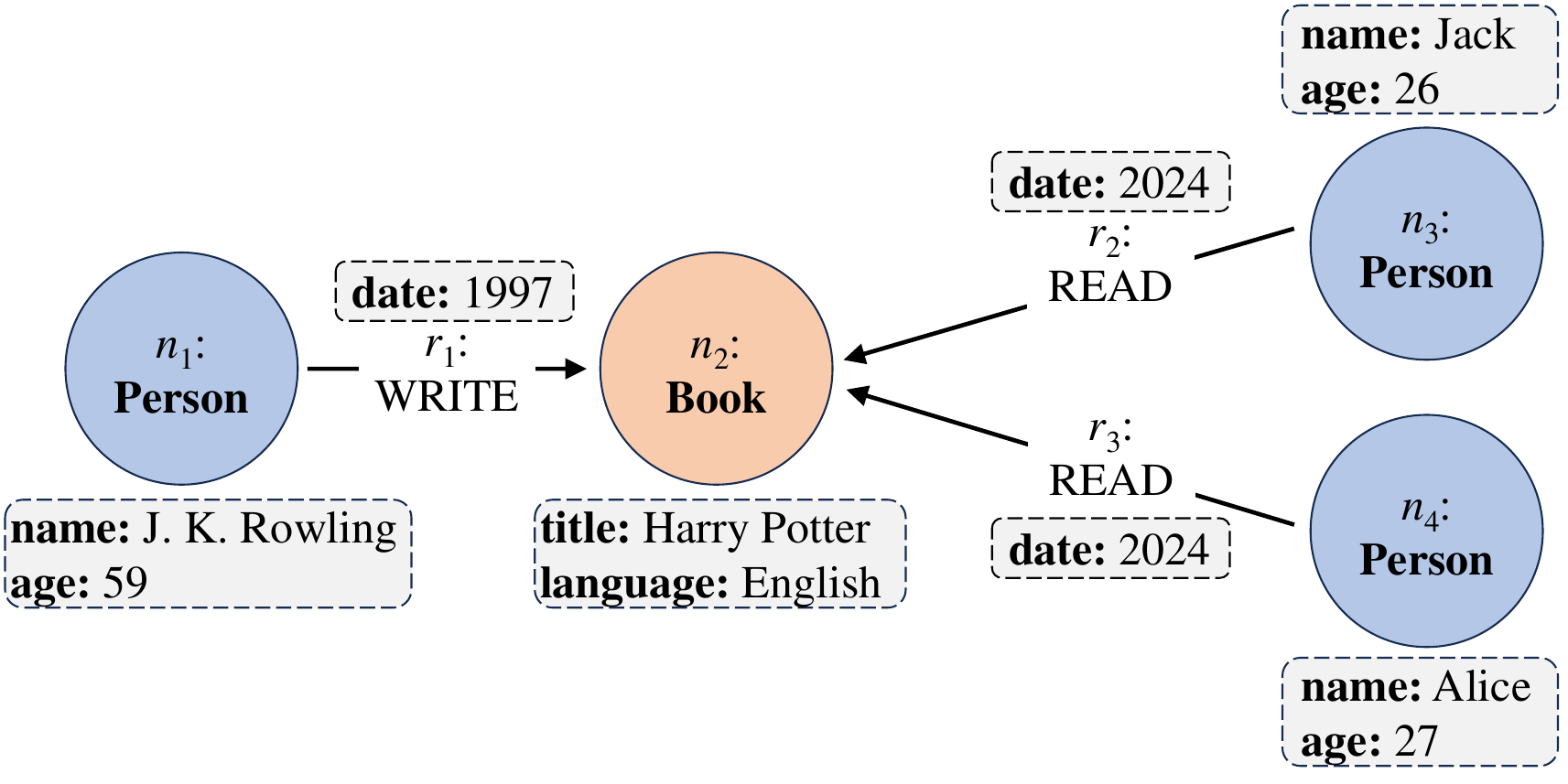}
  \vspace{-3mm}
\caption{An illustrative property graph. We assign a variable
(e.g., $n_1$ and $r_1$) for each node and relationship for easy
reference.}
  \label{fig:property-graph}
\end{figure}

Most GDBs (e.g., Neo4j \cite{kemper2015beginning}, ArangoDB \cite{arangodb} and
OrientDB \cite{ritter2021orientdb}) are built on the property graph model
\cite{angles2017foundations}, in which graph data are stored as nodes and
relationships along with their properties. \figref{fig:property-graph} shows an
illustrative property graph that consists of four nodes and three relationships.
Specifically, each relationship is directed and describes a path from one node
to another (or itself), e.g., relationship $r_1$ connects node $n_1$ and $n_2$.
The property graph model leverages labels to shape the domain of nodes and
relationships. Nodes (or relationships) that have the same label are categorized
into the same set, e.g., node $n_1$, $n_3$ and $n_4$ are categorized by label
$Person$.

% GDBs adopt graph query languages (e.g., Cypher) to define graph patterns, and
% utilize isomorphic \cite{ullmann1976algorithm} or homomorphic
% \cite{barcel2013querying} graph pattern matching to effectively retrieve graph data.
% \listref{list:cypher-example} shows a Cypher graph query that
% defines a graph pattern to retrieve the author of the book read by Alice in the property graph in
% \figref{fig:property-graph}.
% Different from relational database systems that have a standard query language
% SQL, GDBs usually develop their own graph query languages, e.g., Neo4j develops
% Cypher \cite{francis2018Cypher}, Apache TinkerPop \cite{tinkerpop} develops
% Gremlin \cite{gremlin}, and ArangoDB develops AQL \cite{aql}. Among these graph
% query languages, Cypher is one of the most popular graph query languages, and
% has been supported by 4 out of the top 10 GDBs (i.e., Neo4j, Amazon Neptune
% \cite{bradley2018neptune}, Memgraph \cite{memgraph2023} and NebulaGraph
% \cite{wu2022nebula}) in DB-Engines Ranking of GDBs \cite{dbengines2023graph}.

\review{\#1: I found it strange that the paper cites the research paper on Cypher [26] but not the one on GQL (Deutsch et al. Graph Pattern Matching in GQL and SQL/PGQ. SIGMOD 2022), which is much cheaper to access than the standard [31] that's behind a heavy paywall.}

\revision{GDBs adopt graph query languages (e.g.,  Cypher \cite{francis2018Cypher}, Gremlin
\cite{gremlin} and GQL \cite{GQLISO, GQLgraphPattern})}\id{Q11} to define graph
patterns, and utilize isomorphic \cite{ullmann1976algorithm} or homomorphic
\cite{barcel2013querying} graph pattern matching to effectively retrieve graph
data. Among these graph query languages, Cypher is one of the most widely-used
languages, and supported by 4 out of the top 10 GDBs in DB-Engines
Ranking \cite{dbengines2023graph}. \listref{list:cypher-example} shows
an example of a Cypher query that defines a graph pattern to retrieve the
author of the book read by Alice in the property graph in
\figref{fig:property-graph}.

Query equivalence proving is a fundamental problem in database research
\cite{chandra1977optimal, sagiv1980equivalences, cohen1999rewriting}, and has
been widely applied in detecting query optimization bugs
\cite{ganski1987optimization}, mining query rewriting rules
\cite{wang2022wetune},  eliminating query computational overlaps
\cite{zhou2020spes}, etc. We have witnessed the significant progress of SQL
query equivalence provers \cite{green2007krelation, chu2017cosette, chu2018udp,
zhou2019equitas, zhou2020spes, wang2022wetune, ding2023sqlsolver, wang2024QED}
 for relational database systems since 2018.
Initially, syntax-based approaches \cite{green2007krelation, chu2017cosette,
chu2018udp, zhou2019equitas,zhou2020spes} are proposed to prove SQL query
equivalence by checking the syntactic isomorphism of SQL algebraic
representations (e.g., K-relations \cite{green2007krelation}). Although these
approaches can prove SQL query equivalence to a certain extent, they struggle
with handling SQL queries that have significantly different syntactic
structures. Recently,  researchers \cite{wang2022wetune, ding2023sqlsolver,wang2024QED}
propose semantic-based approaches to solve this problem. These approaches use
U-semiring based SQL algebraic representations \cite{chu2018udp} to model the
semantics of SQL queries, and construct the U-semiring expression for a SQL
query (\uexpt{} for short). \uexpt{} is a natural number semiring expression
and models a SQL query $Q$ as an expression $u(t)$ that returns the natural
number multiplicity of a tuple $t$ in $Q$'s query result. Semantic-based
approaches solve the equivalence of SQL queries by proving the equivalence of \uexpt{}s on arbitrary input tuple $t$ based on SMT solvers
\cite{demoura2008z3}. The state-of-the-art semantic-based prover SQLSolver
\cite{ding2023sqlsolver} has demonstrated that its effectiveness significantly surpasses that of syntax-based approaches\cite{chu2018udp, zhou2019equitas,
zhou2020spes}.

\iffalse
\review{Reviewer A: "[a] tabular function [...] cannot model nodes and relationships in property graphs": Why is that? 1) It is not obvious why this should be a true statement. The paper does not provide an argument. 2) SQL/PGQ (the SQL extension for queries property graphs) is doing exactly that, it represents nodes and relationships as tuples in tables.}
\fi

\review{\#Meta Review: Improve the discussion of novelty relative to [16] (R1.O1, R4.O2).}

\review{O1. The novelty wrt [16] should be explained better. For instance, the authors claim that the framework of [16] "is not directly expressive enough" to deal with the graph pattern in Listing 1. But why? This seems like a normal join/project query to me and is, as far as I can see included in the fragment covered by [16].}

\review{\#1: It's a pity that the paper doesn't explain the main new ideas of the approach before diving in details. As I see it, some main new ingredients that this paper needs to deal with compare to [16] are (1) the edge-injective semantics and (2) arbitrary-length paths.}

\review{\#3: O2. The novelty of the approach compared with [16] would require further clarifications.}

\review{\#3: O3. The strength of the contribution and the main technical challenges it addresses compared to SOTA approaches should be clarified further.}

Similar to SQL equivalence provers for relational database systems, graph query
equivalence provers also have significant value for GDBs, and can be applied in
detecting graph query optimization bugs \cite{jiang2023graphgenie,zheng2024dot,zheng2024qudi}, mining graph
query optimization rules \cite{holsch2016graphpattern}, optimizing graph queries
\cite{chandra1977optimal}, etc. However, we still lack a graph query equivalence
prover. \revision{Graph query languages (e.g., Cypher) employ significantly different
syntaxes from SQL, and are built on the property graph model and graph pattern
matching. In contrast, \uexpt{}s \cite{chu2018udp} are built on the relational
data model and tabular tuple calculus utilized by SQL. Therefore, we
cannot directly apply \uexpt{}s to model Cypher
queries.}\id{Q2, Q6, Q12, Q20, Q21}

\revision{To model Cypher queries using \uexpt{}s \cite{chu2018udp}, we first
need to transform property graphs into relational tables. For example, Cytosm
\cite{steer2017cytosm} groups nodes and relationships in a property graph by
their labels and generates an individual table for each label. Then, we can
potentially use \uexpt{}s on these generated tables to model some Cypher
features, e.g., predicates and unions. However, some Cypher features cannot be
modeled in this way due to the limitation of \uexpt{}s. First, based on the
tables generated by Cytosm, \uexpt{}s cannot model node and relationship
patterns either without specifying labels or with multiple labels, e.g., node
pattern \texttt{(writer)} in \listref{list:cypher-example} that is not
associated with any label. Second, \uexpt{}s cannot model some specific Cypher
features, e.g., arbitrary-length paths (e.g., \texttt{()-[*]->()}) and list
variables with \texttt{UNWIND} and \texttt{COLLECT}. Third, Chu et al.
\cite{chu2018udp} cannot support some common features supported by SQL and
Cypher e.g., \texttt{ORDER BY}, \texttt{LIMIT} and \texttt{SKIP}.}\id{Q2, Q6,
Q12, Q20, Q21}

\begin{lstlisting}[float, floatplacement=H,style=cypherstyle, caption={\revision{A Cypher query retrieves the author of the book read by Alice.}\id{Q2, Q6, Q12, Q20}}, label = list:cypher-example, belowskip=-1.5em]
MATCH (reader:Person)-[:READ]->(book:Book)<-[:WRITE]-(writer)
WHERE reader.name = 'Alice'
RETURN writer.name
\end{lstlisting}

\revision{In this paper, we propose a U-semiring based \emph{graph-native} algebraic
representation to accurately model property graphs and Cypher queries,} and
construct U-semiring expressions for Cypher queries (\uexpg{} for short).
\uexpg{}s can model nodes and relationships in property graphs and complex
features in Cypher queries. We then transform proving the equivalence of Cypher
queries into proving the equivalence of \uexpg{}s on an unspecified property
graph. We further leverage the LIA* theory \cite{ding2023sqlsolver} to prove the
equivalence of \uexpg{}s by SMT solvers \cite{demoura2008z3}.

To model property graphs, we define three algebraic functions $Node(e)$,
$Rel(e)$, and $Lab(e, label)$ to precisely model a graph entity $e$'s type
(i.e., node or relationship) and labels (whether $e$ has a label $label$) in
property graphs, respectively. \revision{Thus, \uexpg{}s can model node and
relationship patterns without specifying labels or with multiple labels. For
example, \uexpg{} models the node pattern \texttt{(reader:Person)} as
$Node(reader) \times Lab(reader,Person)$ and the node pattern \texttt{(writer)}
without specifying any label as $Node(writer)$. We further define algebraic
functions to model the outgoing and incoming nodes of a relationship. Based on
these algebraic functions, we construct a U-semiring based algebraic
representation for Cypher's core features, including graph patterns, predicates
and query results, e.g., \texttt{MATCH}, \texttt{WHERE} and \texttt{RETURN}
clauses. We can further model advanced Cypher features, e.g., using
uninterpreted functions in SMT to model arbitrary-length paths, and designing a
divide-and-conquer proving process to handle \texttt{ORDER BY}, \texttt{LIMIT}
and \texttt{SKIP}.} \id{Q2, Q6, Q12, Q20, Q21}

\iffalse
For complex Cypher
queries, e.g., variable-length paths (e.g., \texttt{(n1)-[*1..3]->(n2)}), we
further design a group of Cypher query normalization rules, e.g., eliminating
variable-length paths, to transform them into simplified Cypher queries using
core Cypher features.

\yingying{Remove the following para.}
\tladd{Based on our proposed G-expressions, we then transform proving the
equivalence of Cypher queries into proving the equivalence of \uexpg{}s on an
unspecified property graph. We can further model various Cypher features, e.g.,
using uninterpreted functions in SMT to model arbitrary-length paths, and
designing a divide-and-conquer proving process to handle \texttt{ORDER BY},
\texttt{LIMIT} and \texttt{SKIP}. We further leverage the LIA* theory
\cite{ding2023sqlsolver} to prove the equivalence of \uexpg{}s by SMT solvers
\cite{demoura2008z3}. We also design a group of Cypher query normalization
rules, e.g., eliminating variable-length paths, to transform them into
simplified Cypher queries using core Cypher features.}\id{Q2, Q6, Q20, Q21}
\fi

We implement our approach on Cypher 9 in the openCypher project
\cite{opencypher} as \approachname{}. To evaluate \approachname{}, we
construct a dataset \dataset{} with 148 pairs of equivalent Cypher
queries through the following two methods. (1) We translate the equivalent SQL
query pairs in Calcite \cite{2021calcite}, a widely-used open-source dataset of
equivalent SQL queries
\cite{chu2018udp,zhou2020spes,wang2022wetune,ding2023sqlsolver}, to their
corresponding Cypher queries, and successfully construct 79 pairs of equivalent
Cypher queries. (2) We collect 36 real-world Cypher queries from popular
open-source graph database benchmarks \cite{erling2015ldbc} and widely-used
open-source Cypher projects
\cite{marco2018gdbbenchmark,opencypher_cypher_for_gremlin}. Then, we construct
equivalent Cypher queries by applying three existing Cypher query rewriting
rules \cite{holsch2016graphpattern, jiang2023graphgenie}, and obtain 68 pairs of
equivalent Cypher queries. \dataset{} contains both simple and complex graph
patterns, e.g., optional graph patterns (\texttt{OPTIONAL}) and variable-length
paths (\texttt{-[*2..3]->}). Finally, we evaluate \approachname{} on \dataset{},
and \approachname{} has successfully proven 138 pairs of them with a latency of
38 ms on average. \approachname{} and \dataset{} are available at
https://github.com/choeoe/GraphQe.

In summary, we make the following contributions.
\begin{itemize}[leftmargin=10pt]
\item We propose a U-semiring based \emph{graph-native} algebraic
representation for property graphs and Cypher queries.
\item We propose \approachname{}, the first graph query equivalence prover for Cypher queries.
\item We construct \dataset{}, the first dataset with
148 pairs of equivalent Cypher queries for Cypher query equivalence
proving.
\item We evaluate \approachname{} on \dataset{}. \approachname{} successfully proves 138 out
of 148 pairs of equivalent Cypher queries. 
\end{itemize}
\vspace{-0.25em}

\section{Preliminaries}

In this section, we first introduce the property graph model
(\secref{sec:property-graph-model}). Next, we explain Cypher and its graph
pattern matching mechanism (\secref{sec:graph-pattern-matching}). Finally, we
introduce the U-semiring algebraic structure and how to model SQL semantics
on U-semiring (\secref{sec:U-semiring semantics}).

% \begin{table*}[h]
%   \centering
%   \caption{Core Grammar of Cypher Queries}
%   \vspace{2mm}
%   \renewcommand{\arraystretch}{1.5}
%   \label{tab:cypher-grammar}
%   \begin{tabular}{|ll|}
%   \hline
%   \textbf{Production} & \textbf{Definition} \\
%   \hline
%   \texttt{query} & \texttt{::= RETURN ret | clause query} \\
%   \texttt{ret} & \texttt{::= * | expr (AS a)+ | ret, expr (AS a)+} \\
%   \texttt{clause} & \texttt{::= (OPTIONAL)+ MATCH pattern (WHERE expr)+ | WITH ret (WHERE expr)+ | UNWIND expr AS a} \\
%   \texttt{pattern} & \texttt{::= pattern\_part (, pattern\_part)*} \\
%   \texttt{pattern\_part} & \texttt{::= node\_pattern | node\_pattern rel\_pattern node\_pattern} \\
%   \hline
%   \end{tabular}
% \end{table*}

\subsection{Property Graph Model}
\label{sec:property-graph-model}

Most GDBs (e.g., Neo4j\cite{kemper2015beginning}, JanusGraph
\cite{janusgraph2019}, OrientDB\cite{ritter2021orientdb} and ArangoDB\cite{arangodb}) adopt the property graph model
\cite{kim2012multiplicative, angles2017foundations} to store graph data. In the
property graph model, each node or relationship contains a set of properties.
Each relationship connects a node to another (or itself). The property graph
model utilizes labels to categorize nodes (or relationships) into different
sets, in which all nodes (or relationships) with a certain label belong to the
same set. 

\revision{Different graph query languages may adopt different property graph
models. For example, although GQL is fully compatible with Cypher, they adopt
slightly different property graph models. Cypher only supports directed
relationships and requires each relationship to have only one label, while GQL
further supports undirected relationships and allows a relationship to have more
than one label. Cypher adopts relationship-injective semantics when mapping
relationship patterns to relationships in property graphs, while GQL does not.
In this paper, we mainly focus on Cypher\footnote{GQL is a new standard graph
query language that extends Cypher. Although there are some differences, GQL and
Cypher share many similarities in both syntaxes and semantics. It would be
interesting to perform a similar study for GQL in the future.}, and
we present the formal definition of a property graph model adopted by Cypher as
follows.}\id{Q3, Q8, Q9} 

\review{\#Meta Review: Refine the Formal Treatment addressing the concerns requested by R1.O3, R2.O2.}

\review{\#Meta Review: Discuss the Relationship with GQL, by providing a
detailed comparison between Cypher and GQL, (R1.O4, R4.O1), and aligning the
definition of property graphs in the paper with the standard one proposed by
Francis et al. (ICDT’23).}

\review{\#1: O3. The formal treatment should be less sloppy.}

\review{\#1: In Definition 1, "labels", "key" and "value" are very sloppy. I believe that you need to properly introduce the sets you map from and to. Furthermore, the standard (as in GQL standard) definition of property graphs is different from the one you provide, which is confusing. The definition from the standard is in Francis et al., A Researcher's Digest of GQL, ICDT 2023. In particular, it allows edges to have multiple labels.}

\review{\#2: In Definition 1 $\lambda$ maps to a set of labels, not an individual label. Perhaps making it map into $2^{labels}$ would be better and then restricting the cardinality for relations.}

\revision{
  \begin{definition}
  A property graph used by Cypher is denoted as a tuple $G=\langle N,R,\rho,\lambda,\sigma\rangle$,
  where
  \begin{enumerate}
    \item $N$ is a finite set of nodes used in $G$.
    \item $R$ is a finite set of directed relationships used in $G$.
    \item $\rho: R\rightarrow N\times N$ is a function that maps a directed
    relationship from $R$ to its outgoing and incoming nodes from $N$. 
    \item $\lambda: N\cup R\rightarrow 2^{\mathcal{L}}$ is a labeling function that
   associates each node or relationship to a finite set of labels from
   $\mathcal{L}$ (the full set of labels). In
   Cypher, a node can have one or multiple labels, and a relationship can
   have only one label.\id{Q1, Q9, Q14}
    \item $\sigma: (N\cup R) \times \mathcal{K}\rightarrow Const$ is a partial function
     that associates a constant with a node (or relationship) and a property
     $key$ from $\mathcal{K}$ (the full set of property keys). \id{Q1, Q9}
  \end{enumerate}
\end{definition}
}

\iffalse
\begin{definition} 
  A property graph is denoted as $G=(N,R,\rho,\lambda,\sigma)$,
  where:
  \begin{enumerate}
    \item $N$ is a finite set of nodes.
    \item $R$ is a finite set of relationships.
    \item $\rho: R\rightarrow N\times N$ is a function that maps a relationship
    \tladd{from $R$} to its outgoing and incoming nodes \tladd{from $N$}. 
    \item $\lambda: N\cup R\rightarrow \tladd{2^{labels}}$. \tladd{$labels$ is a
   finite string set containing all the labels in $G$.} $\lambda$ is a function
   that maps a node or relationship to its labels. Note that a node can have one
   or multiple labels, and a relationship can only have one label.
    \item $\sigma: (N\cup R) \times key\rightarrow value$. \tladd{$key$ is string and
     $value$ can be arbitrary type supported by the database system.} $\sigma$ is
     a function that maps a node's (or relationship's) property $key$ to its
     $value$. \id{Q9, Q15}
  \end{enumerate}
\end{definition}
\fi

Take the property graph in \figref{fig:property-graph} as an example. In this
property graph, four nodes form a finite node set $N=$ $\{n_1, n_2, n_3, n_4\}$,
and three relationships form a finite relationship set $R=\{r_1, r_2, r_3\}$.
For relationship $r_1$, which connects the outgoing node $n_1$ and incoming node
$n_2$ (i.e., $\rho(r_1)=\langle n_1,n_2\rangle$), has a label $WRITE$ (i.e.,
$\lambda(r_1)=\{WRITE\}$), and has a property key $date$ whose value is 1997
(i.e., $\sigma(r_1, date)=1997$).

\subsection{Cypher \& Graph Pattern Matching}
\label{sec:graph-pattern-matching}
\begin{figure}[t]
  \centering
  \includegraphics[width=1.0\linewidth]{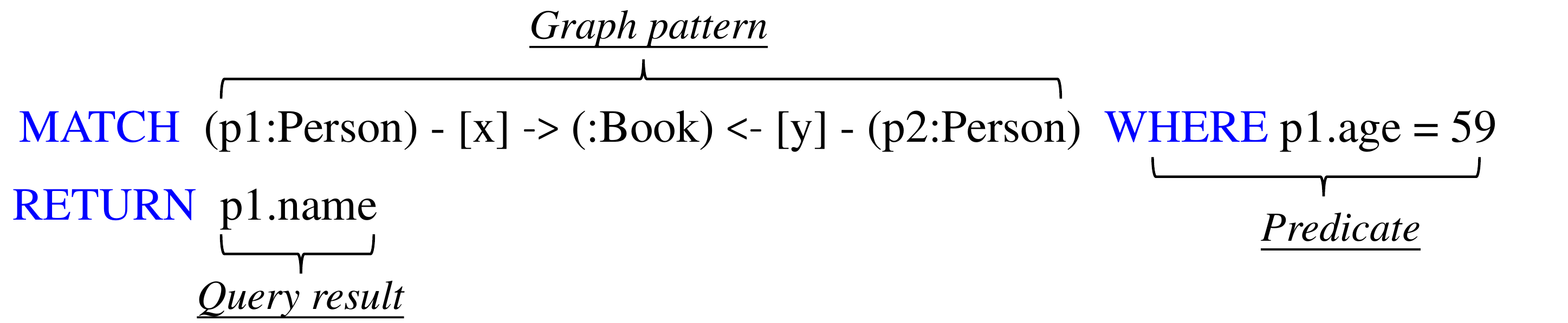}
  \vspace{-6mm}
\caption{The basic structure of a Cypher query.}
\vspace{-4mm}
\label{fig:cypher-query-semantics}
\end{figure}

Cypher is developed by Neo4j \cite{kemper2015beginning} and has been supported
by many GDBs, e.g., MemGraph \cite{memgraph2023}, SAP HANA \cite{farber2012sap},
and NebulaGraph \cite{wu2022nebula}. Cypher is a declarative graph query
language, which adopts graph pattern matching for querying nodes and
relationships in property graphs.

Cypher utilizes \texttt{MATCH} clause to define a graph pattern
(e.g., the graph pattern in \figref{fig:cypher-query-semantics}). This graph pattern
defines three node patterns, i.e., \texttt{(p1:Person)}, \texttt{(:Book)}, and
\texttt{(p2:Person)}, and two relationship patterns, i.e., \texttt{-[x]->}
and \texttt{<-[y]-}. It also contains a predicate \texttt{p1.age=59} specified in the \texttt{WHERE} clause. 
Cypher utilizes the \texttt{RETURN} clause to specify which parts of data should be returned. For example, \texttt{RETURN
p1.name} in \figref{fig:cypher-query-semantics} indicates that the query result
should be the values of the property $name$ of node $p1$.

Besides the above core features, Cypher also supports complex graph patterns.
Cypher allows for multiple graph patterns in chained \texttt{MATCH} clauses,
such as \texttt{MATCH (n1) MATCH (n2)-[]->(n3)}. The results matched by each
graph pattern are combined through Cartesian product. Cypher provides
\texttt{OPTIONAL MATCH} clause to define optional graph patterns that may or may
not exist in a property graph. Cypher graph patterns can include variable-length
paths, e.g., \texttt{(n1)-[*1..3]->(n2)} matches a path with 1 to 3
relationships. Cypher also supports sorting and
aggregates on the query result, such as
\texttt{ORDER BY} or \texttt{COUNT()}. Furthermore, multiple Cypher queries can
be combined by \texttt{UNION ALL} clause that unions the single query
results under bag semantics.

\textbf{Cypher graph pattern matching}. Graph pattern matching serves as the
foundation of graph query languages (e.g., Cypher and Gremlin).
Graph pattern matching maps node and relationship patterns to nodes and relationships in property graph, e.g., node pattern \texttt{(p1:Person)} in the Cypher query of \figref{fig:cypher-query-semantics} can match node $n_1$ in the property graph of \figref{fig:property-graph}. Cypher adopts isomorphism-based graph pattern matching
to query property graphs as follows.

\review{\#1: Definition 2 puzzles me. There is the typo in the definition of $\rho_p$, where U should be x, but phi doesn't seem to be defined at all. (Instead, an example of $\sigma$, which actually doesn't belong in Definition 2, is provided.) Accordingly, item 4 is unclear to me. Furthermore, the edge-injective semantics of Cypher doesn't even occur in Definition 2. It should (and it should be formally defined), because it is a crucial part of the semantics.}

\review{\#2: In Definition 2 there is a missing details regarding labels; namely, that these should be preserved as well, so an item similar to 3. should be added.}

\revision{
  \begin{definition} 
    \label{def:cypher-graph-pattern}
    Given a property graph $G=\langle N,R,\rho,\lambda,\sigma\rangle$, a graph
    pattern on $G$ is defined as $ G_p = \langle N_p,R_p,\rho_p,\phi_p\rangle$.
    Here $N_p$ is a finite set of node patterns, $R_p$ is a finite set of
    relationship patterns, $\rho_p:R_p\rightarrow N_p\times N_p$ is a function
    that projects a relationship pattern to its outgoing and incoming node
    patterns, and $\phi_p$ is a boolean expression constructed on a finite set of
    node patterns from $N_p$ and relationship patterns from $R_p$. A graph pattern
    matching aims to find all possible mappings $f_n$ and $f_r$, which satisfy the
    following conditions. \id{Q1, Q10}
    \begin{enumerate}[label=\arabic*)]
      \item $f_n: N_p\rightarrow N$ maps a node pattern to a node in
       $G$.
      \item $f_r: R_p\rightarrow R$ is an injective mapping that maps a
      relationship pattern to a relationship in $G$.\id{Q1, Q10}
      \item For each $r_n\in R_n$ and $r_p\in R_p$, the specified labels in
      $r_n$ and $r_p$ are a subset of $f_n(n_{p})$ and $f_r(r_{p})$,
      respectively.\id{Q1, Q15}
      \item $f_n$ and $f_r$ are structure-preserving, i.e., for each $r_p\in
      R_p$, if $\rho_p(r_p)=\langle n_{p1}\in N_p, n_{p2}\in N_p\rangle$,
      $f_r(r_p)=r\in R$ and $\rho(r)=\langle n_1,n_2\rangle$, then
      $f_n(n_{p1})=n_1$ and $f_n(n_{p2})=n_2$.
      \item If each node pattern $n_p \in N_p$ and relationship pattern $r_p \in
      R_p$ used in $\phi_p$ are replaced by their corresponding mapped
      node $f_n(n_p)$ and relationship $f_r(r_{p})$ in $G$, $\phi_p$ holds
      True.\id{Q1, Q10}
    \end{enumerate}
  \end{definition}
}

Note that Cypher assigns variables to node and relationship patterns (e.g.,
variable $p1$ in node pattern \texttt{(p1:Person)}) for their references. Node
or relationship patterns that share the same variable are considered to match
the same graph entity. Different from other graph query languages, Cypher
applies injective mapping from relationship patterns to relationships in
property graphs, i.e., they adopt the so-called relationship-injective semantics
\cite{angles2017foundations,kemper2015beginning}. Specifically, in Cypher
queries, different relationship patterns (which are assigned to different
variables) defined within the same \texttt{MATCH} clause are not allowed to
match the same relationship in the property graph. Take the query in
\figref{fig:cypher-query-semantics} as an example. Relationship-injective semantics
require that relationship patterns $x$ and $y$ must map to different
relationships in \figref{fig:property-graph}, e.g.,
\begin{center}
  $f_n(p1)\rightarrow n_1$,
  $f_n(p2)\rightarrow n_3$,
  $f_r(x)\rightarrow r_1$,
  $f_r(y)\rightarrow r_2$
\end{center}
In contrast, relationship-injective semantics do not permit the following
mapping.
\begin{center}
  $f_n(p1)\rightarrow n_1$,
  $f_n(p2)\rightarrow n_1$,
  $f_r(x)\rightarrow r_1$,
  $f_r(y)\rightarrow r_1$
\end{center}

\subsection{U-semiring \& U-semiring SQL Semantics}
\label{sec:U-semiring semantics}
\iffalse
\tladd{U-semiring is an algebraic structured on natural number. Many SQL query
equivalence provers \cite{chu2018udp,wang2022wetune,ding2023sqlsolver} adopt
U-semiring to model SQL queries into algebraic representations for equivalence
proving. In this section, we will present the standard model of U-semiring and
how it models a SQL query for automatically equivalence proving.}
\fi

\textbf{U-semiring}. Unbounded semiring (U-semiring for short) \cite{chu2018udp}
is a natural number semiring with unbounded summation. U-semiring extends the
commutative semiring \cite{green2007krelation} with new operators ($\sum,
\|\cdot\|, not(\cdot)$) and is defined as follows.

\begin{definition}
  U-semiring is a commutative natural number semiring denoted as
  $(\mathbb{N},0,1,+,\times,\|\cdot\|, not(\cdot), \sum)$, where
  \begin{enumerate}
    \item The squash operator $\|\cdot\|$ is unary and transforms an input value
    into an output between 0 and 1, e.g., $\|0\|=0$, $\|1+x\|=1$. Squash
    operator is commonly used to deduplicate query results under bag semantics.
    \item The $not(\cdot)$ operator is unary and reverses an input boolean value,
    e,g, $not(1)=0$, $not(0)=1$.
    \item $\sum_{t\in D}E(t)$ takes an expression $E(t)$ as input and
    outputs a natural value. Specifically, $\sum$ is used to enumerate all
    variable (set) $t$ within a given domain $D$ for $E(t)$ and accumulate the output
    values.
  \end{enumerate}
\end{definition}

U-semiring also adopts semiring operator $[\phi]$ from K-relation \cite{green2007krelation} that
takes a boolean expression $\phi$ as input, and outputs 1 for true, 0 for false.
$[\phi]$ transforms a boolean value into integer for arithmetic $\times$ and $+$
under semiring semantics.

\textbf{U-semiring SQL algebraic representation}. Based on U-semiring, Chu et al. \cite{chu2018udp} propose an algebraic representation to model SQL
queries under bag semantics and construct
U-semiring expressions for SQL (\uexpt{} for short).
\uexpt{} defines table function $R(t)$ to return the multiplicity of tuple $t$ in table
$R$. Then \uexpt{} models a SQL query as an expression $u(t)$ that returns the
multiplicity of tuple $t$ in the query result. For example, a SQL query
\begin{center}
  \texttt{SELECT c1 FROM R WHERE c2 = 1}
\end{center}
is modeled as a \uexpt{}
\vspace{-4pt}

{\footnotesize\begin{center}
$u(t)=\sum_{t_1}[t=t_1.c1]\times R(t_1)\times
[t_1.c2=1]$
\end{center}}

\vspace{-4pt}
This \uexpt{} returns the multiplicity of an arbitrary tuple $t$ in the query
result by counting all possible tuple $t_1$ in table
$R$ using $\sum_{t_1}$, and filters the predicate $t_1.c2=1$ using $[t_1.c2=1]$.
$[t=t_1.c1]$ expresses the projection that each tuple $t$ in the query result is
the column $c1$ of a tuple $t_1$ in $R$. 

\uexpt{} can also model complex SQL features. For subqueries combined by
\texttt{UNION}, \uexpt{} recursively models the subqueries and combines them
using semiring operator $+$. For tables joined together, \uexpt{} multiplies
table functions, e.g., $R(t_1)\times S(t_2)$.

By using \uexpt{}s, proving the equivalence of SQL queries is transformed
into proving the equivalence of \uexpt{}s. To prove the equivalence of \uexpt{}s,
UDP \cite{chu2018udp} formalizes \uexpt{}s using
axioms, creating a canonical form that allows for
isomorphism checking between \uexpt{}s. Additionally, SQLSolver
\cite{ding2023sqlsolver} leverages LIA* theory to eliminate $\sum$ in \uexpt{}s,
modeling them as first-order logical expressions that are verified by SMT
solvers.

\section{Overview}

To prove the equivalence of Cypher queries, we first formalize the problem of
Cypher query equivalence under bag semantics in
\secref{sec:overview-formalizing}. We then explain the basic idea of modeling
the Cypher semantics based on U-semiring in \secref{sec:overview-our-approach}.
Finally, we introduce the workflow of our Cypher equivalence prover in
\secref{sec:overview-workflow}.

\subsection{Problem Formulation}
\label{sec:overview-formalizing}
% \tl{What is Cypher QE --> multiplicity of Cypher tabular result tuple --> how
% the multiplicity is calculated equivalent --> how to express it in \uexpg{}}
\iffalse
Before we define a formulation of Cypher query equivalence, we present the
features of the Cypher query tabular result under bag semantics including the
result tuple and its multiplicity.

\begin{itemize} [leftmargin=10pt]
  \item \textbf{Cypher tabular result tuple}. Cypher returns tabular
  results under bag semantics, in which columns are defined by the
  \texttt{RETURN} clause. Cypher tabular result can be sorted using
  \texttt{ORDER BY} clause and automatically grouped by aggregate functions.

  \item \textbf{Cypher result tuple multiplicity}. Under bag semantics, Cypher
  allows for duplicated tuples in the result set. Therefore we use the
  multiplicity of tuples to model Cypher query result. The multiplicity of a
  Cypher result tuple is the number it appears in the Cypher query result set.
\end{itemize}
\fi

Cypher queries return tabular results under bag semantics that allow for
duplicates. Two Cypher queries $Q_1$ and $Q_2$ are equivalent if their results
consist of the same tuples, and the multiplicity of any tuple $t$ in their
results is equal. Additionally, if Cypher queries return ordered results (e.g.,
queries with \texttt{ORDER BY} clauses), any two tuples $t_1$ and $t_2$ appear in
the same order in the results of both queries $Q_1$ and $Q_2$. 

For two Cypher queries $Q_1$ and $Q_2$ with \texttt{ORDER BY} clauses, if their
corresponding sub-queries $Q_{1}^{'}$ and $Q_{2}^{'}$ without \texttt{ORDER BY} clauses
are equivalent, and $Q_1$ and $Q_2$ sort the query results returned by
$Q_{1}^{'}$ and $Q_{2}^{'}$ according to the same \texttt{ORDER BY} expressions, we can say
that $Q_1$ and $Q_2$ are equivalent.

Therefore, the Cypher query equivalence under the ordered bag semantics can be
defined as follows.

\begin{definition}
  \label{def:cypher-equality-2}
  If two Cypher queries $Q_1$ and $Q_2$ are equivalent, they need to satisfy the
 following conditions.
  \begin{enumerate}
    \item The results of $Q_1$ and $Q_2$ contain the same tuples.
    \item The multiplicity of any tuple $t$ in the results of $Q_1$ and $Q_2$ is
equal.
    \item If $Q_1$ or $Q_2$ contains the \texttt{ORDER BY} clauses, their corresponding
    sub-queries $Q_{1}^{'}$ and $Q_{2}^{'}$ without \texttt{ORDER BY} clauses are
    equivalent, and $Q_1$ and $Q_2$ sort the query results returned by
    $Q_{1}^{'}$ and $Q_{2}^{'}$ according to the equivalent \texttt{ORDER BY} expressions.
  \end{enumerate}
\end{definition}

\subsection{Basic Idea}
\label{sec:overview-our-approach}

\iffalse
In this paper, we propose a U-semiring based Cypher algebraic representation to
model the semantics of Cypher queries, and construct U-semiring expressions for
Cypher queries (\uexpg{} for short). \uexpg{} can model nodes and relationships
in property graphs and complex features in Cypher queries. We then transform
proving the equivalence of Cypher queries into proving the equivalence of
\uexpg{} on any property graph. We further leverage the LIA* theory
\cite{ding2023sqlsolver} to prove the equivalence of \uexpg{} by SMT solver
\cite{demoura2008z3}.

To model property graphs, we define three \emph{boolean} algebraic functions
$Node(e)$, $Rel(e)$, and $Label(e)$ to express a graph entity $e$'s types (i.e.,
node or relationship) and labels (whether $e$ has a label $Label$) in property
graphs, respectively. We further define algebraic functions to express the
incoming and outgoing nodes of a relationship. With these algebraic functions,
we construct a U-semiring based algebraic representation for Cypher's core
features, including graph patterns and query results, e.g., e.g.,
\texttt{MATCH}, \texttt{WHERE} and \texttt{RETURN} clauses in Cypher queries.
For complex Cypher queries, e.g., variable-length paths (e.g.,
\texttt{(n1)-[*1..3]->(n2)}) and built-in functions (e.g., $toLower()$), we
design a group of Cypher query normalization rules, e.g., eliminating
variable-length paths, and translate them into
\tlreplace{simplified}{de-sugared} Cypher queries using core Cypher features.
\fi

Inspired by the \uexpt{}s for modeling SQL queries, we propose an approach to
model a Cypher query as a \uexpg{} $g(t)$,
which takes a tuple $t$ in the Cypher query result as input and outputs its
multiplicity. Then two Cypher queries $Q_1$ and $Q_2$ are considered
semantically equivalent if their corresponding \uexpg{}s $g_1(t)$
and $g_2(t)$ always output the same value for any tuple $t$ on an unspecified property
graph.

To model a Cypher query based on U-semiring, we first assign a variable $e_i$ to
represent any property graph entity for each node pattern and relationship
pattern in a graph pattern $G_p = \langle N_p, R_p, \rho_p, \phi_p \rangle$. We
then model the Cypher query as a \uexpg{} based on these variables. Finally, we
can calculate the multiplicity of a tuple $t$ in the Cypher
query result by enumerating all possible graph entities ($e_1, \dots, e_n$) over
an unspecified property graph $G$, which satisfy  graph pattern $G_p$.

\review{\#1: The provided example on page 4 is somewhat disappointing because (1) I think that it doesn't illustrate a difference with [16] apart from Cypher syntax and (2) g(t) is always 0 or 1, which means that it doesn't illustrate bag semantics.}

For example, given a simple Cypher query \texttt{MATCH (n1)-[r]->(n2)
WHERE n1.age=59 RETURN n1}, we
% \begin{center}
%   \begin{minipage}{\textwidth}
%     \texttt{MATCH (n1)-[r]->(n2) WHERE n1.age=59} \\
%     \texttt{RETURN n1}
%   \end{minipage}
% \end{center}
assign three variables $e_1$, $e_2$, $e_3$ to node pattern \texttt{(n1)},
relationship pattern \texttt{-[r]->}, and node pattern \texttt{(n2)},
respectively. Then, we model the Cypher query into a \uexpg{} as 
\vspace{-7pt}

{\footnotesize\begin{align*}
g(t) = &
\sum_{e_1, e_2, e_3}[t=e_1] \times Node(e_1) \times Rel(e_2) \times Node(e_3)\\&
  \times [in(e_2)=e_1]\times [out(e_2)=e_3] \times [e_1.age=59]
\end{align*}}

\noindent in which $Node(e)$, $Rel(e)$, $in(e)$ and $out(e)$ denote whether $e$ is a node
entity, relationship entity, $e$'s incoming node and outgoing node,
respectively. \revision{ $g(t)$ uses the unbounded summation ($\Sigma$) to
enumerate all graph entities ($e_1$, $e_2$, $e_3$) over an unspecified property
graph $G$ and calculates the multiplicity of tuple $t$ in the Cypher query
result using $[t=e_1]$.}\id{Q12}

\revision{Note that $g(t)$ returns a natural number for any tuple $t$, and can
illustrate the bag semantics of Cypher queries.}\id{Q12} Based on this idea, we
further construct \uexpg{}s for the predicates and sorting of a
Cypher query and a set of complex Cypher features, e.g., the arbitrary-length
paths (\texttt{(n1)-[*]->(n2)}). The problem of proving Cypher query equivalence
is then transformed into that of proving the equivalence of their corresponding
\uexpg{}s, i.e., $g_1(t)$ and $g_2(t)$.

\subsection{Workflow of \approachname{}} 
\label{sec:overview-workflow}

\begin{figure}[t]
  \centering
  \includegraphics[width=0.8\linewidth]{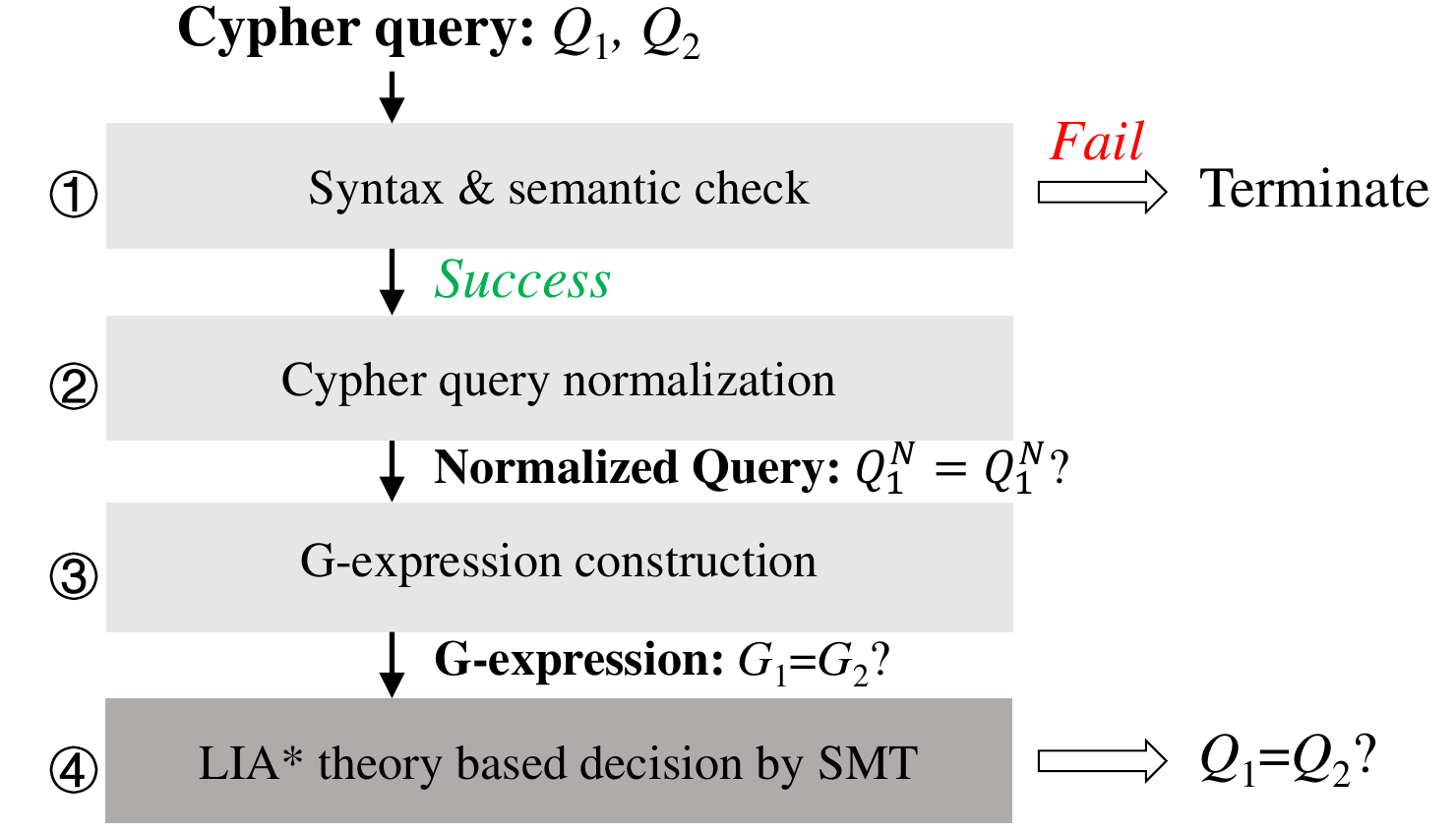}
\caption{The  workflow of \approachname{}.}
  \vspace{-4mm}
  \label{fig:workflow}
\end{figure}

We propose an approach to model Cypher queries based on our idea in
\secref{sec:overview-our-approach} and implement it as \approachname{}.
\approachname{} takes a pair of Cypher queries $Q_1$ and $Q_2$ as input and
returns whether they are equivalent. As shown in
\figref{fig:workflow}, \approachname{} consists of
four stages.
\iffalse
\review{Reviewer A: It is perfectly fine to limit this work to some subset of Cypher. However, it should be clear early on (like Sec 1), what we are looking at.}
\fi

\textbf{ \textcircled{1} Syntax \& semantic check}. The input Cypher queries may
contain syntax or semantic errors. Syntax errors occur when Cypher queries
violate the Cypher grammar and semantic errors occur when Cypher queries use
illegal operations. To ensure the correctness of the input Cypher queries, we
perform both syntax and semantic checks, discarding any queries that contain
errors. Specifically, for a given Cypher query, we first attempt to construct an
Abstract Syntax Tree (AST) using a Cypher grammar parser that is built on the
openCypher grammar \cite{opencypher}. Queries that fail to generate ASTs are
considered to violate the Cypher grammar, and the proving process is terminated.
For syntactically correct Cypher queries, we adopt a semantic check based on
their ASTs. We check the following cases: (1) Incorrect variable reference.
Undefined variable references in \texttt{WHERE} clause cause semantic errors.
(2) Incorrect relationship labels. In property graphs, each relationship can
have only one label. Relationship patterns that share the same variable but
define different labels can lead to semantic errors. Similar to syntax checking,
for queries that fail the semantic check, we terminate the proving process. 
\iffalse
\review{Reviewer A: This would be an excellent candidate for experimental validation. How much does the normalization step reduce the proving latency? Is it for any kind of query a net positive?}
\fi

\textbf{ \textcircled{2} Cypher query normalization}. Since some complex Cypher
queries (e.g., those with variable-length paths) are difficult to be directly
modeled and some Cypher queries can be isomorphic after simple query transforms,
we apply a rule-based Cypher query normalization to simplify these
queries into forms that we can support. Specifically, we
define a group of normalization rules on the Cypher ASTs to handle built-in
functions, complex features and reform the ASTs. We will further discuss our
rule-based Cypher query normalizations in \secref{sec:normalization}.

\textbf{\textcircled{3} \uexpg{} construction}. For
normalized Cypher queries $Q_1^N$ and $Q_2^N$, we model them as U-semiring Cypher
expressions (\uexpg{} for short) $G_1=g_1(t)$ and $G_2=g_2(t)$, which model the
multiplicity of tuple $t$ in their query results. By utilizing \uexpg{}s, we transform
the problem of proving Cypher query equivalence into proving the equivalence of
$g_1(t)$ and $g_2(t)$. We will further discuss our \uexpg{}s in \secref{sec:U-semiring-Cypher-semantics}.

\textbf{\textcircled{4} Decision procedure}. Once Cypher queries are modeled into \uexpg{}s,
we prove the equivalence of \uexpg{}s by proving that $\exists t. g_1(t)\neq
g_2(t)$ is unsatisfiable using the Z3 SMT solver \cite{demoura2008z3}. Since SMT
solvers cannot handle the unbounded summations (i.e., $\sum$) in \uexpg{}s, we
leverage the LIA* based algorithm proposed by Ding et al. \cite{ding2023sqlsolver}
to eliminate the unbounded summations and construct first-order logical
expressions for proving their satisfiability by Z3. 

\iffalse
\begin{table}[t]
  \centering
  \caption{The algebraic functions in \uexpg{} for modeling property graphs and
  Cypher semantics.}
  \label{tab:extended_u_expression} 
  \begin{tabular}{|m{2cm}|m{2cm}|m{3.4cm}|}
  \hline
  \textbf{Function description} & \textbf{Function definition} &
  \textbf{Function values} \\
  \hline
  Node qualifier & $Node(t)$ & If $t$ is a node, return 1;
  otherwise, 0 \\
  \hline
  Relationship qualifier & $Rel(t)$ & If $t$ is a
  relationship, return 1; otherwise, 0 \\
  \hline
  Label qualifier & $Label_{i}(t)$ & If $t$ has label $label$, return 1;
  otherwise, 0 \\
  \hline
  Incoming node projector  & $in(t)\times Rel(t)$ & Returns the incoming node of relationship
  $t$ \\
  \hline
  Outgoing node projector & $out(t)\times Rel(t)$ & Returns the outgoing node of
  relationship $t$ \\
  \hline
  Sorting \& grouping & $order(t)$, $group(t)$& Tag result tuple $t$ to express sorting and grouping\\
  \hline
  Get column & $Get_i(t)$ & Get the $i$-th column of result tuple $t$\\
  \hline
  \end{tabular}
\end{table}
\fi

\section{G-expression}
\label{sec:U-semiring-Cypher-semantics}
To model Cypher queries as \uexpg{}s, we first model the property graph model
under semiring semantics (\secref{sec:modeling-property-graph}). Then we model
core and advanced Cypher features, and construct \uexpg{}s for Cypher queries
(\secref{sec:model-cypher-query}). Finally, we illustrate the decision procedure
of proving the equivalence of \uexpg{}s (\secref{sec:prove-uexpr-equiv}).

\iffalse
\begin{figure}[t]
  \centering
  \includegraphics[width=1.0\linewidth]{t-entities.pdf}
  \caption{The structure of tagged graph entities (node and relationship). We
  define tagged relationships with an ordered pair of outgoing and incoming
  nodes for modeling property graph structures.}
  \label{fig:relational_graph}
\end{figure}
\fi

\subsection{Modeling Property Graph} 
\label{sec:modeling-property-graph}
\iffalse
Property graph entities are stored as nodes and relationships under set
semantics. Therefore, we do not model the multiplicity of each property graph
entity, but represent whether it is a node or a relationship. Specifically, we
define $Node(e)$ and $Rel(e)$ functions to model each property graph $e$.

\textbf{Node function} checks if a property graph entity is a node. For a
property graph entity $e$ in a property graph $G=(N,R,\rho,\lambda,\sigma)$:
\[
Node(e) = \begin{cases} 
1 & \text{if }e\in N \\
0 & \text{if }e\in R 
\end{cases}
\]

\textbf{Relationship function} checks if a property graph entity is a
relationship. For a property graph entity $e$ in a property graph
$G=(N,R,\rho,\lambda,\sigma)$:
\[
Rel(e) = \begin{cases} 
1 & \text{if }e\in R \\
0 & \text{if }e\in N 
\end{cases}
\]
\fi

Property graph entities are stored as nodes and relationships under set
semantics, so that the multiplicity of each property graph entity is one.
Therefore, we only need to distinguish whether a property graph entity is a node
or a relationship. For an arbitrary property graph entity $e$ in an unspecified
property graph $G=\langle N,R,\rho,\lambda,\sigma\rangle$, we define $Node(e)$
and $Rel(e)$ functions to model nodes and relationships as follows.
\vspace{-4mm}

{\[Node: N\cup R\rightarrow \mathbb{B}=\{1,0\},\text{  } Rel: N\cup R\rightarrow
\mathbb{B}=\{1,0\}\]
\vspace{-4mm}
\begin{align*}
  Node(e) &= \begin{cases} 
    1 & \text{if } e \in N \\
    0 & \text{otherwise} 
   \end{cases} &
   Rel(e) &= \begin{cases} 
    1 & \text{if } e \in R \\
    0 & \text{otherwise} 
   \end{cases}
\end{align*}}

\vspace{0mm}
Each relationship in a property graph connects two nodes from its outgoing node
to its incoming node. 
For a relationship $e$ in an unspecified property graph
$G=\langle N,R,\rho,\lambda,\sigma\rangle$, we define $out(e)$ and $in(e)$ functions to express a
relationship $e$'s outgoing and incoming nodes, respectively, as follows.
\[out: R\rightarrow N,\text{  }in: R\rightarrow N\]
 Specifically, if $e_1$ and $e_2$ are the outgoing and incoming
nodes of $e$, respectively, i.e., $\rho(e)=\langle e_1, e_2\rangle$, then
\[out(e) = e_1,\text{  } in(e) = e_2\] To express the multiplicity of relationships'
outgoing and incoming nodes, we use U-semiring operator $[\cdot]$ on
$out(e)$ and $in(e)$. For example, $[out(e)=e_1]$ returns 1 if $e_1$ is the
outgoing node of relationship $e$, and 0 otherwise. Note that, $out(e)$ and
$in(e)$ are always used together with $Rel(e)$ to ensure $e$ is a complete relationship
in a property graph.

Property graphs utilize labels to categorize nodes (relationships) into
different node (relationship) sets. To model labels of each property graph
entity, we define a function $Lab(e, label)$ to represent that a property graph
entity $e$ has a label $label$. Specifically, for an arbitrary graph entity $e$
in an unspecified property graph
$G=\langle N,R,\rho,\lambda,\sigma\rangle$, we
define $Lab(e, label)$ as follows.
\vspace{-4mm}

\[
Lab: (N \cup R) \times \mathcal{L} \rightarrow \mathbb{B} = \{1,0\}
\]

\vspace{-4pt}
Specifically, for the labels of $e$, denoted as $\lambda(e)$, we have
\vspace{-4pt}

\[
Lab(e, label) = 
\begin{cases} 
1 & \text{if } label \in \lambda(e) \\
0 & \text{otherwise}
\end{cases}
\]

\vspace{-3pt}
Property graph entities have a set of properties. We model each property of a
property graph entity by defining $x.key$ to access the property value of a given
property name $key$. For example, we use $e.p_1$ to access the value
of $e$'s property name $p_1$, and $[e.p_1=v_1]$ returns 1 if $v_1$ is the value
of $e.p_1$ and 0 otherwise.

\subsection{Modeling Cypher Query Features} 
\label{sec:model-cypher-query}
\review{\#1: I went through Section IV.B, which shows how Cypher features are mapped to algebraic expressions, but found the approach to be a bit underwhelming.}

Based on the predefined functions in \secref{sec:modeling-property-graph}, we
model Cypher queries as U-semiring based Cypher expressions (\uexpg{} for
short). \revision{\figref{fig:cypher-fragments} shows the Cypher fragments we support currently.}\id{Q4, Q7} We first model core features of Cypher queries, e.g.,
graph patterns, predicates and query results. We then
model a set of advanced Cypher features, e.g., join of graph patterns.

\review{\#Meta Review: Define and describe a Clear Fragment of Cypher which
the translation works and clearly state the scope of the claims (R1.O2).}

\review{\#1: O2. It is unclear for which fragment of queries the approach works. The paper would be much stronger if a soundness and completeness result for a natural fragment of queries would be provided. (Much like Figure 2 in [16].)}

\begin{figure}[t]\footnotesize
  \centering
  \revision{
  \begin{equation*}
    \begin{aligned}
      q \in \text{Query} ::&= m\; r\; 
       | \; q_1 \text{ UNION } q_2\;|\; q_1 \text{ UNION ALL } q_2 \\
      m \in \text{Match} ::&= \text{MATCH } p_1,\ldots,p_n\\
        &| \; \text{OPTIONAL MATCH } p_1,\ldots,p_n\\
        &| \; m \; \text{WHERE}\; b\;
        | \; m \;\text{WITH} \; b_1,\ldots b_n \\
      r \in \text{Return} ::&= b\; | \; \text{DISTINCT} \; r\;
        | \; r \; \text{ORDER BY} \; b \; \\
        &| \; r \;  \text{LIMIT} \; b \; 
        | \; r \; \text{SKIP} \; b\\
      p \in \text{GraphPattern} ::&= n
        \; | \; n\; c\\
      n \in \text{Node}::&= \text{(}v:l_1:...:l_n\; \{a_1...a_n\}\text{)} \\
      v \in \text{Variable}::&= \text{string} \\
      l \in \text{Label}::&= \text{string} \\
      a \in \text{Property}::&= \text{string} \\
      c \in \text{Relationship} ::&= p\text{-[}v:l_1:...:l_n \; \{a_1...a_n\} *i_1...i_2 \text{]-}\texttt{>}n \\
        &| \; p\texttt{<}\text{-[}v:l_1:...:l_n \; \{a_1...a_n\} *i_1...i_2 \text{]-}n \\
        &| \; p\text{-[}v:l_1:...:l_n \; \{a_1...a_n\} *i_1...i_2 \text{]-}n \\
      i \in \text{Constant} ::&= \text{integer}\\
      b \in \text{Expression} ::&= b_1 = b_2\;|\; b_1\neq b_2\\
        &| \; \text{NOT } b\;|\;b\text{ IS NULL }|\; +b\;|\; -b\;|\; (b)\\ 
        &| \; b_1 \text{ AND } b_2 \; | \; b_1 \text{ OR } b_2 \\
        &| \; b_1 > b_2 \; | \; b_1 < b_2 \;|\;b_1 \geq b_2 \; | \; b_1 \leq b_2 \\
        &| \; \text{TRUE} \; | \; \text{FALSE} \; | \; \text{EXISTS } q \\
        &| \; v\;|\; v.a \; | \; f(b) \; | \; \text{agg}(b) \\
        &| \; * \; | \; b \text{ AS } s \; | \; p_1, p_2 \;| \; i\;
        |\; \text{UNWIND}\; b \\
      agg ::&=\text{COLLECT}\;|\;\text{COUNT}\;|\;\text{SUM}\;|\;\text{MAX}\;
        |\;\text{MIN}\; |\; \text{AVG}\\
        s \in \text{Alias} ::&= \text{string}\\
        f \in \text{UDF} ::&= \text{string}
  \end{aligned}
  \end{equation*}
  }
  \vspace{-4mm}
  \caption{\revision{Cypher fragments supported by
  \approachname{}.}\id{Q4, Q7}}
  \vspace{-4mm}
  \label{fig:cypher-fragments}
\end{figure}

\subsubsection{Modeling the Core Features of Cypher Queries}
\review{\#3: D1. Does the G-expression representation introduce computational overhead when processing highly complex queries?}

Cypher provides several core features to form a basic query for graph
pattern matching, including graph patterns, predicates, and query
results, as shwon in \figref{fig:cypher-query-semantics}.

\textbf{Modeling Cypher graph patterns}. Cypher graph patterns include node and
relationship patterns. We model each node or relationship pattern as an
algebraic expression, which outputs 1, if an arbitrary graph entity $e$
satisfies this pattern; otherwise outputs 0.

A complete node pattern in a Cypher query consists of a variable that refers to
this node pattern, labels and properties with their values. To model a node
pattern, we assign an arbitrary graph entity $e$ to it and construct an
algebraic expression using predefined functions. Specifically, we use $Node(e)$
to represent that property graph element $e$ has the type of node, use
$Lab(e,label)$ to represent $e$ has label $label$, and use $[e.key=value]$ to
represent the value of its property $key$ is $value$. Furthermore, we use AND
(i.e., $\times$) operations to join multiple labels and multiple properties. For
example, a node pattern \texttt{(n:l1:l2\{p1:v1\})} is assigned a variable $n$,
and has two labels $l1$ and $l2$ and a property $p1$ with its value $v1$. We can
model this node pattern as $Node(e)\times Lab(e,l1)\times Lab(e,l2)\times
[e.p1=v1]$. Note that the variable (e.g., \texttt{n}), the labels (e.g.,
\texttt{:l1:l2}), and the properties (e.g., \texttt{\{p1:v1\}}) can be omitted
in a node pattern. For example, $()$ is the simplest node pattern, and we only
use $Node(e)$ to model it.

A complete relationship pattern in a Cypher query consists of a variable that
refers to this relationship pattern, labels and properties with their values,
and its outgoing and incoming node patterns. To model a relationship pattern, we
assign an arbitrary graph entity $e$ to it and construct an algebraic expression
using predefined functions. Specifically, we use $Rel(e)$ to represent that
property graph pattern $e$ has the type of relationship, use $Lab(e,label)$ to
represent $e$ has label $label$, use $[e.key=value]$ to represent the value of
its property $key$ is $value$, and use $[out(e) = e_1]$ and $[in(e)=e_2]$ to
represent $e$'s outgoing node pattern is $e_1$ and incoming node pattern is
$e_2$, respectively. For multiple properties, we use AND (i.e., $\times$)
operations to join them. However, different from node patterns, for multiple
labels in a relationship pattern, we use OR (i.e., $+$) operations instead of
AND (i.e., $\times$) operations to connect them, because a relationship
can have only one label. For example, a relationship pattern
\texttt{(n1){-}[r:l1|l2\{p1:v1\}]{->}(n2)} is assigned a variable $r$, has two
labels $l1$ and $l2$, a property $p1$ with its value $v1$, and its outgoing
and incoming node patterns are assigned variables $n1$ and $n2$, respectively. We can
model this relationship pattern as $Rel(r)\times[out(r)=n2]\times
[in(r)=n1]\times (Lab(r, l1)+Lab(r, l2))\times [r.p1=v1]$. 

Cypher adopts relationship-injective semantics for the relationship patterns
that require relationship patterns with different variables to match different
relationships in the property graph. To model it, we construct not equal
predicates to each pair of relationships defined within the same \texttt{MATCH}
clause. For example, for a graph pattern \texttt{MATCH ()-[r1]->()-[r2]->()}, we
first assign $e_1$ and $e_2$ to relationship pattern \texttt{()-[r1]->()} and
\texttt{()-[r2]->()}, respectively. Then we construct expression
$not([e_1=e_2])$ for restricting that the two relationship patterns must match
the different relationships. Note that relationships defined in different
\texttt{MATCH} clauses are not restricted by relationship-injective semantics,
e.g., $r1$ and $r2$ in \texttt{MATCH ()-[r1]->() MATCH ()-[r2]->()}.

\textbf{Modeling predicates}. Predicates define filtering conditions in
a Cypher query. We leverage the semiring operator $[\cdot]$ to model predicates.
For example, we model the predicate in
\figref{fig:cypher-query-semantics} using $[e.age=59]$, which returns 1 if the property $age$ of entity $e$ is 59,
and 0 otherwise. Multiple predicates are connected by $\times$ for
\texttt{AND} operations and $+$ for
\texttt{OR} operations. For example, we model \texttt{WHERE
p1.age>29 OR p1.age<59} under semiring semantics as
$\|[e_1.age>29]+[e_1.age<59]\|$. 

\iffalse
\begin{definition}
  $in(e)$ or $out(e)$ takes a tagged relationship $e$ as input and output its
  incoming or outgoing node, where:
  \begin{enumerate}
    \item $in(e)$: If $e$ is a tagged relationship with node
    sequence $<N_1,N_2>$, output its incoming node $N_1$.
    \item $out(e)$: If $e$ is a tagged relationship with node
    sequence $<N_1,N_2>$, output its outgoing node $N_2$.
    \item $in(e)$ and $out(e)$ are always used together with $Rel(e)$ to
    indicate that $e$ is a tagged relationship.
  \end{enumerate}
\end{definition}

With the algebraic functions above, \uexpg{} models the data entities and graph
structure of a Cypher graph pattern by constructing a projection for each
relationship to its incoming and outgoing nodes. Besides the graph data entities
and graph structures, Cypher defines graph patterns with conditions. Conditions
on a graph pattern can modeled with U-semiring operator $[\cdot]$ that
takes an expression as input and output 1 for true, e.g., \texttt{WHERE
n1.age=59} is modeled as $[n1.age=59]$ under U-semiring semantics.
\fi
\begin{table*}[h!]
  \caption{Modeling advanced Cypher features by \uexpg{}s.}
  \vspace{-5pt}
  \label{tab:all-summary}
  \centering
  \begin{tabular}{|l|l|l|}
  \hline
  \textbf{Cypher feature} & \textbf{Query example} & \textbf{\uexpg{}}\\ 
  \hline
  Intermediate results &\makecell[l]{\texttt{MATCH (n) WITH n.name AS name}\\ 
  \texttt{RETURN name}} & $\sum_t'([t=t']\times \sum_{t'}([t'=e.name]\times
  Node(e)))$\\
  \hline
  Aggregate& \texttt{MATCH (n:Person) RETURN SUM(n.age)} &
  \makecell[l]{$g(t)=[t=\sum_{t'}[t'=e_1.name]\times Node(e_1)\times
  Lab(e_1,Person)$\\ $\times e_1.sal$}\\
  \hline
  Unwinding &\makecell[l]{\texttt{WITH [\{c1:0, c2:1\},\{c1:2, c2:3\}]}\\
  \texttt{AS tmp UNWIND tmp AS tmpRow}} &
  \makecell[l]{$([tmpRow.c1=0]\times [tmpRow.c2=1])$\\ $+([tmpRow.c1=2]\times
  [tmpRow.c2=3])$}\\
  \hline
  \iffalse
  Existence &\texttt{WHERE ()-[]->()} & \makecell[l]{$\|\sum_{e_1,e_2,e_3}(Node(e_1)\times
  Rel(e_2)\times$\\$Node(e_3)\times [out(e_2)=e_3]\times [in(e_2)=e_1])\|$} \\
  \hline
  \fi
 Arbitrary-length path &\texttt{()-[*]->()} &
  $Rel(e)\times UNBOUNDED(e)\times [out(e) = e_1]\times [in(e)=e_2]$ \\
  \hline
  Sorting with truncation &\makecell[l]{\texttt{WITH x.name} \texttt{ORDER BY
  x.age}\\ \makecell[l]{\texttt{RETURN 1}}}& $[t=e.name]\times
  [asc(t)=e.age]\times [limit(t)=1]$\\
  \hline
  Natural join & \texttt{MATCH...}$(q_1)$  \texttt{MATCH...}$(q_2)$  &
  \makecell[l]{$G(q_1)\times G(q_2)$} \\
  \hline
  Left outer join & \makecell[l]{\texttt{MATCH...} $(q_1)$  \texttt{OPTIONAL MATCH...}}$(q_2)$ &
    \makecell[l]{$G(q_1)\times G(q_2) + G(q_1)\times not(G(q_2))\times isNULL(G(q_2))$} \\
  \hline
  Union all & \makecell[l]{\texttt{MATCH...} $(q_1)$ \texttt{UNION ALL MATCH...} $(q_2)$} & $G(q_1)+G(q_2)$ \\
  \hline
  Union & \makecell[l]{\texttt{MATCH...} $(q_1)$ \texttt{UNION MATCH...} $(q_2)$} &
    $\|G(q_1)+G(q_2)\|$ \\
  \hline
  \iffalse
  Dedup & \texttt{RETURN DISTINCT x }& $\| \sum([t=e]\times\dots) \|$ \\
  \hline
  \fi
  \end{tabular}
  \vspace{-10pt}
\end{table*}

\textbf{Modeling query results}. Cypher queries return tabular results that
consist of a set of tuples, each of which has a set of
column values. We model each tuple in the result table using $t.col_i$ and
construct the projections from graph entities to the value of $t.col_i$. For
example, in \figref{fig:cypher-query-semantics}, the tuple in the query result
has one column, which projects an arbitrary graph entity $e_1$ for node pattern
$p1$ to the value of its property $name$. Thus, we model this query result using
$[t.col_1=e_1.name]$. 

\subsubsection{Modeling the Advanced Cypher Features} 
\label{sec:uexp-construction}
\review{\#2: In Section IV-B.2 (pages 7,8) there is a misuse of edge labels, where relations are written as e.g. $Person(e_1)$ instead of $Rel(e_1)$ AND $Label(e_1,Person)$. This inconsistency in writing is present throughout the discussion on nested subqueries.}

Cypher is highly expressive and provides a set of advanced features or clauses
to enrich its expressiveness. As shown in \tabref{tab:all-summary}, we model a
set of advanced Cypher features and construct \uexpg{}s for them. 

\review{\#3: D3. How robust is GraphQE when dealing with queries that produce extremely large intermediate results?}

\textbf{Intermediate results}. The \texttt{WITH} clause creates intermediate
results in a query. It generates an intermediate table as the input of the
subsequent clause. For each intermediate table, we introduce an intermediate
variable \( t'_i \) to model the multiplicity of tuples in the table in the same
way as \texttt{RETURN}. For Cypher queries with the structure: \texttt{MATCH}
$p_1$ \texttt{WITH} $i_1, i_2...$ \texttt{MATCH} $p_2$ \texttt{RETURN} $e_1,
e_2...$, we model it as $g(t)=\sum\nolimits_{t', \vec{e_1}}(E_2\times
[t.col_1=e_1]\times...\times \sum\nolimits_{\vec{e_2}}(E_1\times
[t'.col_1=i_1]\times...))$, where \( E_1 \) represents the \uexpg{} for \( p_1
\), \( E_2 \) represents the \uexpg{} for \( p_2 \), and \( t' \) is the
intermediate variable. Furthermore, temporary variables are also used to model
Cypher nested subqueries, such as \texttt{MATCH (n) WHERE EXISTS { MATCH (n1)
WHERE n1.p1 > 100 } RETURN n}.

However, temporary variables may lead to semantic loss. For example, for the
Cypher query
\texttt{MATCH (p) WITH DISTINCT p.name} \texttt{AS name RETURN
name},
we construct the following \uexpg{}s for it.
\vspace{-6pt}

{\footnotesize\begin{align*}
g(t)=&\sum_{t'}[t = t']\times \|\sum_{e_1}Node(e_1) \times [t' = e_1.name]\|
\end{align*}}

\vspace{-6pt}
\noindent The \uexpg{}s cannot be correctly solved by SMT solvers since $t'$ lost the
uniqueness of the query results created by \texttt{WITH DISTINCT p.name}.
Therefore, we propose a normalization rule to eliminate temporary variables.
Specifically, we remove temporary variable $t'$ from the unbounded summation if
it can be represented by other variables using $[t'=t_1]$. After normalization,
$Q$'s \uexpg{} becomes
\vspace{-6pt}

{\footnotesize\begin{align*}
\revision{g(t)=\|\sum_{e_1}Node(e_1) \times Lab(e_1, Person) \times [t= e_1.name]\|}\id{Q16}
\end{align*}}

\iffalse
with no temporary variables left in the predicates. However, for another
possible \uexpg{} $g(t)=\sum_{t'}(\|\sum_{e_1}(Node(e_1) \times Lab(e_1, Person)
\times [t'.col_1=e_1.name]\times [t'.col_2=e_1.age])\|\times [t = t'.col_1])$,
this rule can not remove $t'$.
\fi 
\vspace{-6pt}
\noindent Intermediate variables can cause predicates to deviate from their intended
scopes, e.g., the $[e_1.dept = e_2.dept]$ in 
\vspace{-6pt}

{\footnotesize\begin{align*}
g(t)=&\sum_{e_1}[t = e_1] \times Node(e_1) \times
[e_1.dept = e_2.dept]\times \| \sum_{e_2}Node(e_2)\|
\end{align*}}

\vspace{-6pt}
\noindent To address this, we propose a normalization rule that moves the
predicate to the correct summation body as follows.
\vspace{-6pt}

{\footnotesize\begin{align*} g(t)=&\sum_{e_1}[t = e_1] \times Node(e_1)\times
  \| \sum_{e_2}Node(e_2)\times [e_1.dept = e_2.dept]\| \end{align*}}

\vspace{-6pt}
\textbf{Aggregate}. \approachname{} models aggregates using intermediate
variables. For example, for aggregate \texttt{SUM} in the Cypher query
segment \texttt{MATCH (n:Person) RETURN SUM(n.age)}, \approachname{} creates
variable $t'$ and model it as 
\vspace{-6pt}

{ \footnotesize
  \begin{align*}
g(t)=[t=\sum_{t'}[t'=e_1.name]\times Node(e_1)\times Lab(e_1,Person)\times e_1.sal]
\end{align*}
}

\vspace{-4pt}
\approachname{} does not model the concrete semantics of \texttt{COLLECT} but
represents it using an uninterpreted function in the SMT solver.

\iffalse
\review{Reviewer A: Cypher is not limited to unwind a constant list. The list is given by an expression.}
\fi

\textbf{Unwinding}. The \texttt{UNWIND} clause transforms a list into individual
rows. The list for unwinding can be a constant list created by \texttt{WITH} or
collected from \texttt{COLLECT}. As shown in \tabref{tab:all-summary}, we model
unwinding on a constant list by modeling the concatenation of each element in it
using $+$. Unwinding a collected list will break it into individual rows and
remove the aggregates, e.g., \texttt{UNWIND(COLLECT(n.name))} is directly
modeled into \texttt{n.name}.  

\iffalse
\textbf{Returning collected list}. Cypher uses \texttt{COLLECT()} clause to
create list, in which the rows are defined inside the clause. We model
\texttt{COLLECT()} in the \texttt{RETURN} clause as function $collect(t,i)$ on
the columns of arbitrary tuple in the query result, which outputs 1 if the
$i$-th column of $t$ is collected and 0 otherwise. For example, we model
\texttt{RETURN COLLECT(x1.name)} as $[t.col_1=e_1.name]\times collect(t,1)$.

\textbf{Pattern existence}. Cypher declares graph patterns in the \texttt{WHERE}
clause to specify their existence, e.g., \texttt{WHERE ()-[]->()} requires there
exists such a graph pattern \texttt{()-[]->()} in the property graph. We
use the semiring operation
$\|\cdot\|$ to express the existence of the given pattern. For example, we model
\texttt{WHERE ()-[]->()} as $\|\sum_{e_1,e_2,e_3}(Node(e_1)\times
Node(e_3)\times
Rel(e_2) \times [in(e_2) = e_3] \times [out(e_2) = e_1])\|$. Besides, we also
use $\|\cdot\|$ to model \textbf{dedup} that is often denoted as
\texttt{DISTINCT} in Cypher. For example, given a Cypher query \texttt{MATCH (n) RETURN
DISTINCT n}, we model it as $\|\sum_{e}[t=e]Node(e)\|$.
\fi

\review{\#2: While arbitrary reachability patterns are supported (e.g. -[*]-$>$), what about the ones with a specified label set (e.g. -[:a*]-$>$). It seems that the proposed approach would not work in this case. It would be good to have a discussion on this. I understand that full regular expressions are probably outside of the scope (also not fully supported in Cypher), but the query that just uses one edge label or a set of labels could be incorporated into the UNBOUNDED operator probably.}

\textbf{Arbitrary-length path}. Cypher uses \texttt{(n1)-[*]->(n2)} to retrieve
all paths of any length between two nodes $n1$ and $n2$. In this case, $n1$ and $n2$
are fixed and the relationships in the paths must satisfy the same predicate.
Therefore, we treat arbitrary-length path patterns (i.e., \texttt{()-[*]->()}) as a
special kind of relationship pattern, and assign a relationship variable $e$ to
this pattern, in which $e_1$ and $e_2$ represent its outgoing and incoming node
patterns, respectively. Then, we define a function $UNBOUNDED(e)$ on $e$ that
outputs 1 if $e$ is a combination of any number of relationships. Finally, we
construct \uexpg{} term $Rel(e)\times UNBOUNDED(e)\times [out(e) = e_1]\times
[in(e)=e_2]$ to model this pattern. \revision{Besides, labels specified on the
arbitrary-length path can also be supported through this approach. For example,
we model \texttt{-[*:KNOWS]->} as $UNBOUNDED(e)\times Lab(e, KNOWS)$.} \id{Q16, Q17}

\iffalse
\review{Reviewer A: This is not a safe assumption w.r.t. to language semantics. It is not clearly specified for openCypher either way. It may be observable behavior in many Cypher engines that do not involve intra query parallelism. However, the Neo4j documentation says:
"When ORDER BY is present in a WITH clause, the immediately following clause
will receive records in the specified order. The order is not guaranteed to be
retained after the following clause, unless that also has an ORDER BY
subclause." From a language semantics point of view, this makes sense since it
gives room for more intra query parallelism.}
\review{Reviewer A: Cypher has also SKIP. The paper leaves the reader guessing that SKIP is modeled analogous to LIMIT.}
\fi

\textbf{Sorting with truncation}. Cypher sorts and truncates the query results
using the \texttt{ORDER BY...LIMIT...SKIP...} fragments. To model \texttt{ORDER BY
$o_1,...,o_n$}, we follow an idea that a condition on the query result can be
represented as the same condition on all the tuples within the query result.
Therefore, we treat the $o_1,...,o_n$ as conditions on all the tuples in the
query result. We define function $order(t,i)$ on an arbitrary tuple $t$ in the
query result to represent $o_i$. Then, we construct $[order(t,i)=o_i]$ to model
the value of $o_i$. Finally, we use $\times$ to connect all the
$[order(t,i)=o_i]$, i.e., $[order(t,1)=o_1] \times \dots
\times[order(t,n)=o_n]$. To model \texttt{LIMIT }$l$ and \texttt{SKIP }$s$, we
also treat them as conditions on all the tuples in the query result. We define
functions $limit(t)$ and $skip(t)$ that represent the limiting and skipping
condition on all the tuple $t$ in the Cypher query result. Then, we model
\texttt{LIMIT }$l$ and \texttt{SKIP }$s$ as $[limit(t)=l]$ and $[skip(t)=s]$. 

For the \texttt{ORDER BY...LIMIT...SKIP...} fragments among Cypher subqueries,
we consider the following cases: (1) single \texttt{ORDER BY} is ignored since
the order it specifies will not be guaranteed by the following clauses. (2) For
\texttt{ORDER BY} followed by \texttt{LIMIT} and \texttt{SKIP}, we cannot
directly model it into a single \uexpg{}. Instead, we design a
divide-and-conquer based approach that individually check the equivalence of
their subqueries. For example, for the equivalent Cypher queries in
\listref{list:inner-sorting}, we divide each query into subqueries, i.e., $Q1$
into $Q1'$ and $Q1''$, $Q2$ into $Q2'$ and $Q2''$. Then we check the equivalence
of each pair of subqueries, i.e., $\langle Q1', Q2'\rangle$, and $\langle Q1'', Q2''\rangle$. \\
\vspace{-10pt}
\begin{lstlisting}[style=cypherstyle,label={list:inner-sorting}, caption={Equivalent Cypher queries with \texttt{ORDER BY...LIMIT...}\\ within their subqueries.}]
Q1:   MATCH (n1) WITH n1 ORDER BY n1.p1 LIMIT 1  
      MATCH (n1)-[]->(n2) RETURN n2
Q1':  MATCH (n1) WITH n1 ORDER BY n1.p1 LIMIT 1
Q1'': MATCH (n1) MATCH (n1)-[]->(n2) RETURN n2

Q2:   MATCH (n1) WITH n1 ORDER BY n1.p1 LIMIT 1 
      MATCH (n2)<-[]-(n1) RETURN n2
Q2':  MATCH (n1) WITH n1 ORDER BY n1.p1 LIMIT 1
Q2'': MATCH (n1) MATCH (n2)<-[]-(n1) RETURN n2
\end{lstlisting}

\textbf{Product/Concatenation.} Cypher graph patterns defined in multiple
\texttt{MATCH} are joined through Cartesian product, and we recursively
construct \uexpg{}s (denoted as $G(s_i)$ in \tabref{tab:all-summary}) for each
\texttt{MATCH} and connect them using $\times$. Cypher graph patterns are left
outer joined by \texttt{OPTIONAL MATCH}, and we model the graph patterns that
can be NULL using uninterpreted function $isNULL$ in the SMT solver. In the
similar way, we use $+$ for the concatenation of recursively constructed
\uexpg{}s of Cypher subqueries in \texttt{UNION ALL} and use an extra
$||\cdot||$ for \texttt{UNION}.

\subsection{Proving the Equivalence of \uexpg{}}
\label{sec:prove-uexpr-equiv}

\iffalse
\review{Reviewer C: It is not clear what LIA* theory is. More information is
required (just a few more lines) not to leave the reader wandering on what this
is. This part with LIA* formulas should be somehow streamlined. How the LIA*
formula is constructed?
}
\fi

By constructing \uexpg{}s for Cypher queries, we transform the problem of proving
the equivalence of Cypher queries into proving the equivalence of \uexpg{}s. We
first map the returned elements across the two Cypher queries. This process
prevents query inequality caused by difference in the order of returned
elements. For example, consider two Cypher queries as follows.\\
$Q_1$: \texttt{MATCH (n1)-[r]->(n2) RETURN n1, n2}\\
$Q_2$: \texttt{MATCH (n1)<-[r]-(n2) RETURN n1, n2}\\ In
$Q_1$, the returned element $n1$ should be mapped to the returned element $n2$
in $Q_2$, and should not be affected by their orders. To achieve this, we pair
the returned elements in both queries according to their types. Specifically,
node or relationship variables are mapped to those of the same type, expressions
are mapped to expressions of the same type, and references to variable
properties are mapped to references with the same name for variables of the same
type. If no successful mappings are found, we will remove these conditions and
map again. If the numbers of elements returned by the two Cypher queries are
inconsistent, then the two queries can only be equivalent if they both return
empty results on any property graph. Therefore, we directly prove whether the
two queries satisfy this case.

To prove the equivalence of \uexpg{} $g_1(t)$ and $g_2(t)$,
we prove that $\exists t. g_1(t)\neq g_2(t)$ is unsatisfiable through Z3 SMT
solver. Since SMT solvers cannot handle unbounded summations, we leverage LIA*
theory based algorithm \cite{ding2023sqlsolver, levatich2020solving} to eliminate unbounded
summations. LIA* formula extends Linear Integer Arithmetic to represent
unbounded summations, allowing for reasoning unbounded summations using SMT
solvers. The algorithm replaces each unbounded summations with an integer value
$v_i$ and finds an equisatisfiable formula to represent the unbounded summations.

For example, given two \uexpg{}s
\vspace{-6pt}

{\footnotesize\begin{align*}
  g_1(t)=&\sum_{e_1}[t=e_1.name] \times Node(e_1)\times([e_1.age<10]\\ &+[e_1.age>20])\\
  g_2(t)=&\sum_{e_1}[t=e_1.name] \times Node(e_1)\times [e_1.age<10]\\ &+\sum_{e_2}[t=e_2.name] \times Node(e_2)\times [e_2.age>20]
\end{align*}}

\vspace{-6pt}
\noindent the algorithm replaces the unbounded summations in $g_1(t)$ and $g_2(t)$ and
models them as  $v_1$ and $v_2+v_3$. It finds an equisatisfiable formula of
$\exists v_1, v_2, v_3. v_1\neq v_2+v_3\wedge
(v_1,v_2,v_3)=\lambda_1(1,0,1)+\lambda_2(0,1,0)$, which can be proved
unsatisfiable by SMT solvers since such $\lambda_1$ and $\lambda_2$ do not
exist.

\begin{table*}[t]
  \centering
  \renewcommand{\arraystretch}{1}
  \caption{Normalization rules used for de-sugaring complex Cypher queries.}
  \label{tab:normalized}
  \resizebox{\textwidth}{!}{
    \begin{tabular}{|c|m{3.5cm}|m{6cm}|m{5.6cm}|}
      \hline
      \textbf{No.} & \textbf{Normalization rule} & \textbf{Original query} & \textbf{Normalized query} \\
      \hline
      \textcircled{1} & Eliminating undirected relationship pattern &
      \texttt{MATCH (n1)-[]-(n2) RETURN n1.name} &
      \makecell[l]{\texttt{MATCH (n1)-[]->(n2) RETURN n1.name}\\\texttt{UNION ALL}\\\texttt{MATCH (n1)<-[]-(n2) RETURN n1.name}} \\
      \hline
      \textcircled{2} & Rewriting variable-length path &
      \texttt{MATCH (n1)-[*1..2]->(n2) RETURN n1} &
      \makecell[l]{\texttt{MATCH (n1)-[]->(n2) RETURN n1}\\\texttt{UNION ALL}\\\texttt{MATCH (n1)-[]->()-[]->(n2)}\\ \texttt{RETURN n1}} \\
      \hline
      \textcircled{3} & Rewriting \texttt{RETURN *} &
      \texttt{MATCH (x)-[z]->()-[y]->() RETURN *} &
      \texttt{MATCH (x)-[z]->()-[y]->() RETURN x, y, z} \\
      \hline
      \textcircled{4} & Eliminating redundant clause &
      \texttt{MATCH (x) WITH x.name AS name RETURN name} &
      \texttt{MATCH (x) RETURN x.name} \\
      \hline
      \textcircled{5} & Standardizing variable &
      \texttt{MATCH (person)-[]->(book) RETURN person} &
      \texttt{MATCH (n1)-[r1]->(n2) RETURN n2} \\
      \hline
      \textcircled{6} & ID equality simplification &
      \texttt{MATCH (n1), (n2) WHERE id(n1)=id(n2) RETURN n2} &
      \texttt{MATCH (n1) RETURN n1} \\
      \hline
    \end{tabular}
  }
  \vspace{-10pt}
\end{table*}

\section{Rule-based Cypher Query Normalization}
\label{sec:normalization}

\iffalse
\begin{algorithm}[t]
  \caption{Procedure for normalizing a Cypher query.}
  \label{alg:normalize_cypher}
  \renewcommand{\algorithmicrequire}{\textbf{Input:}}
  \renewcommand{\algorithmicensure}{\textbf{Output:}}
  \begin{algorithmic}[1]
  \Require Cypher query AST $Q$
  \Ensure Normalized Cypher query AST $Q'$
  \State $Q' \gets Q$
  \State $Q' \gets string\_function\_replacement(Q')$
  \State $Q'\gets eliminating\_undirected\_relationships(Q')$
  \State $Q'\gets variable\_length\_path\_rewriting(Q')$
  \State $Q'\gets return\_rewriting(Q')$
  \State $Q'\gets removing\_as\_clauses(Q')$
  \State $Q'\gets standardizing\_references(Q')$
  \State $Q'\gets id\_equality\_simplification(Q')$
  \State \Return $Q'$
  \end{algorithmic}
  \end{algorithm}
\fi

Cypher provides some complex features that cannot be directly modeled by the
approach in \secref{sec:U-semiring-Cypher-semantics} (e.g., variable-length
paths (\texttt{()-[*1..2]->()}) and \texttt{RETURN *}. However, we observe that
these features can be represented by the combinations of features in
\secref{sec:U-semiring-Cypher-semantics}. Therefore, we propose a group of
normalization rules that transform queries containing these complex features
into equivalent queries using only the features we have modeled. Specifically,
each normalization rule traverses the Cypher AST, matches specific query
fragments, and transforms them into simplified equivalent Cypher fragments.

% \textbf{Variable linking}. We search for equivalent expressions in the Cypher
% query statement. If one side of the equivalent expression is a variable, we
% construct a link from the variable to the other side of the expression. If both
% sides are variables, the order of linking does not affect the subsequent
% processing results. Later, we can simplify the construction of algebraic
% representations by eliminating additional variables through direct handling of
% the expressions corresponding to variables via linking.

\iffalse
\subsection{\tladd{Cypher Query} Normalization\tldelete{ Rules}}
\fi

\label{sec:normalization-rules}
\tabref{tab:normalized} shows the normalization rules to transform complex
features. Rule \textcircled{1} transforms an undirected relationship in a Cypher
query to the union of relationships with both directions. This is because Cypher
does not support querying undirected edges. Instead, undirected edges are parsed
as two directed edges in opposite directions. Rule \textcircled{2} transforms
variable-length paths into the union of all the lengths. Rule
\textcircled{3} to \textcircled{5} aims to fill the omitted but determined parts
of a Cypher query, e.g., \texttt{RETURN *}. Rule \textcircled{6} converts
primary key equivalences into variable equivalences based on the integrity
constraints of the graph database.

We normalize a Cypher query by sequentially applying these normalization rules
round by round until no rule is successfully applied. Only one rule is applied
per round to avoid conflicts. Specifically, \textcircled{5} is applied after
\textcircled{2}, \textcircled{3} and \textcircled{4}, since \textcircled{2} and
\textcircled{4} can create anonymous node or relationship patterns and copy the
existing patterns that should be standardized by \textcircled{5}. Since
\textcircled{5} assign variables to anonymous graph patterns, it is applied
after \textcircled{3}. We apply \textcircled{6} after \textcircled{5}, because
we should not replace the variable names of node or relationship patterns across
subqueries since they are indeed different.

\section{Soundness and Completeness}

\review{\#Meta Review: Formally state and include a proof sketch that
the translation of a Cypher query Q to an equivalent arithmetic expression $g_Q$
produces the same results for simple queries, as requested by R2.O1.}

\review{\#2: O1. While it is rather obvious that transforming a Cypher query Q to an equivalent arithmetic expression $g_Q$ does in fact produce the same results for simple queries, perhaps it would be good to have this formally stated as a theorem and have a proof sketch somewhere in the text.O2. There are some minor technical inconsistencies with the writing (details below), but nothing serious.
O3. Some query features are not covered in full detail.}

\review{\#2: I would recommend adding a theorem that states the soundness of the transformation of a query Q into $g_Q$. Namely that the result set is the same. A proof sketch would be nice as well. If space is the issue I would suggest dropping Section 8.}

\iffalse
\review{Reviewer B: The paper provides an incomplete algorithm for Cypher query equivalence, but we do not know whether the problem is decidable at all (and if not, for which fragment of the language it would be). We also do not have a precise characterization of the cases in which the proposed algorithm may fail to detect equivalence of two queries. \\
I’m missing here a complexity analysis, which definitely should be provided.
Without it we cannot understand on which parameters the efficiency depends on
and to correctly understand performance evaluation later in Section 7.4\\
What about deciding on query containment? I believe a discussion is missing as
well on this rather important problem.}
\fi

\textbf{Soundness}. We now show that \approachname{}, our
approach for checking the equivalence of Cypher queries, is sound.

\revision{\emph{Theorem 1}. Let $Q$ be a simple Cypher query, which is defined
by the Cypher fragments shown in \figref{fig:cypher-fragments}, excluding
arbitrary-length path, built-in functions, \texttt{ORDER BY}, \texttt{LIMIT},
and \texttt{SKIP}. }\id{Q5, Q13} Let $g(t)$ be the \uexpg{} representation of
$Q$. Then, for any tuple $t$, $g(t)$ returns the multiplicity of $t$ in the
evaluation of $Q$ over a property graph $G$ under Cypher bag semantics.

\revision{\emph{Proof sketch}. According to Definition 2, Cypher queries
adopt graph pattern matching that finds all maps from each node/relationship
pattern ($N_p\cup R_p$ in $G_p$) to structure-preserving node/relationship in a
property graph ($N\cup R$ in $G$) and satisfy condition $\phi_p$. \uexpg{}
$g(t)$ uses algebraic functions and semiring operations to translate each Cypher
feature in $Q$ into natural number semiring semantics, which computes the
multiplicity of each tuple $t$ by counting all the combinations of property
graph elements matched in $G$. For example, given a simple Cypher query $Q$:
\texttt{MATCH (n1)-[r]->(n2) RETURN n1}, \approachname{} maps the graph
pattern in $Q$ into the product of algebraic functions as $Node(n1)\times
Rel(r)\times Node(n2)\times [in(r)=n1]\times [out(r)=n2]$ on semiring semantics
that is structure preserving. Then, \approachname{} maps the graph pattern matching in
$Q$ into $\sum_{n1,r,n2}$ that enumerates all the combinations of property graph
elements and calculates the multiplicity of tuple $t$ in $Q$'s query result
using $[t=n1]$. Other simple Cypher features are modeled in the similar way. Therefore,
$g(t)$ returns the multiplicity of any tuple $t$ in $Q$'s query result.
$\square$}\id{Q5, Q13}

\emph{Theorem 2}. Given two Cypher queries $Q_1$ and $Q_2$ and their
corresponding translated \uexpg{}s $g_1(t)$ and $g_2(t)$, if $g_1(t)$ and
$g_2(t)$ are equivalent, then $Q_1$ and $Q_2$ are equivalent.

  \emph{Proof sketch}. We prove this theorem by examining three different cases.
  \begin{enumerate}[leftmargin=15pt]
    \item $Q_1$ and $Q_2$ do not contain arbitrary-length paths,
    built-in functions, sorting and truncation. According to Theorem 1,
    $g_{1}(t)$ and $g_{2}(t)$ return the multiplicities of arbitrary tuple $t$
    in their query results. If the SMT solver proves $\exists t. g_{1}(t)\neq
    g_{2}(t)$ is unsatisfiable, then for an arbitrary tuple $t$, the
    multiplicities of $t$ in the query results of $Q_1$ and $Q_2$ are the same.
    Thus, $Q_1$ and $Q_2$ are equivalent.
    \item $Q_1$ and $Q_2$ contain arbitrary-length paths, built-in functions,
    sorting and truncation outside subqueries. \approachname{} models these
    features as uninterpreted functions in $g_1(t)$ and $g_2(t)$. The
    equivalence of $g_1(t)$ and $g_2(t)$ implies that $Q_1$ and $Q_2$  use these
    features in the same way. This implies that the equivalence of $g_1(t)$ and
    $g_2(t)$ is the sufficient condition for the equivalence of $Q_1$ and $Q_2$.
    \item $Q_1$ and $Q_2$ contain sorting and truncation within subqueries.
    \approachname{} divides $Q_1$ and $Q_2$ into subqueries $Q_1^1\dots Q_1^m$
    and $Q_2^1\dots Q_2^n$. Then, \approachname{} requires $m=n$ and individually
    proves each $Q_1^i$ and $Q_2^i$ are equivalent, which forms a sufficient
    condition for the equivalence of $Q_1$ and $Q_2$. $\square$
  \end{enumerate}

\textbf{Completeness}. \approachname{} does not ensure completeness. That
said, even if two Cypher queries are equivalent, we cannot ensure their corresponding
translated $g_1(t)$ and $g_2(t)$ are equivalent. The incompleteness of
\approachname{} is caused by the following reasons. (1) \approachname{} does not
model all Cypher features, e.g., regular functions and \texttt{CALL}. (2)
\approachname{} does not support handling all cases for \texttt{ORDER
BY...LIMIT...SKIP...} fragments in subqueries. Our divide-and-conquer based
approach forms a sufficient but not necessary condition proving the equivalence
of Cypher queries. (3) \approachname{} utilizes uninterpreted functions to model
arbitrary-length paths and built-in functions, which only form a sufficient but
not necessary condition for proving the equivalence of Cypher queries. (4)
\approachname{} relies on the algorithm proposed by Ding et al
\cite{ding2023sqlsolver} to prove the equivalence of \uexpg{}s that is also
incomplete.

\section{Evaluation}
\vspace{-0.3mm}
\iffalse
\review{Reviewer B: The evaluation of the algorithm is limited, but this is
justified by the fact that no query sets were available, and in fact the paper
provides a first benchmark set for query equivalence. The set should include
also query pairs that are not equivalent, but differ only minimally, to check
that the algorithm is indeed sound, and does not classify two queries as
equivalent, when they are not.\\
Reviewer C: Complexity analysis is completely missing.
Evaluation seems weak, without any baselines.}
\fi

We implement \approachname{} with around 5000 lines of Java code. We use Antlr4
\cite{parr2013antlr4} to parse Cypher queries into Cypher Abstract Syntax Trees
(ASTs) according to the grammar from openCypher \cite{opencypher}. We then
transform the ASTs into graph relational algebra
\cite{holsch2016graphpattern,marton2017formalising}. We construct a \uexpg{} for
each Cypher query based on its graph relational algebra. The decision procedure
is implemented based on the LIA* construction algorithm in SQLSolver
\cite{ding2023sqlsolver} and utilizes Microsoft Z3 \cite{demoura2008z3} to
verify the equivalence of \uexpg{}s.

To demonstrate the effectiveness of \approachname{}, we address the
following two research questions:

\begin{itemize}[leftmargin = 12pt]
    \item \textbf{RQ1:} How effective is {\approachname{}}'s proving capability?
    \item \textbf{RQ2:} How is the performance of \approachname{} on proving
    Cypher query equivalence?
\end{itemize}

To answer \textbf{RQ1}, we construct a dataset of 148 equivalent Cypher query
pairs as \dataset{} and test the verification capability of \approachname{} on
it. To the best of our knowledge, \approachname{} is the first equivalence
prover for Cypher queries. Therefore, we cannot compare \approachname{} with
other provers. However, we analyze the verification result of \approachname{} on
all the Cypher features in \dataset{}.

To answer \textbf{RQ2}, we calculate the verification latency of \approachname{}
on each test case in \dataset{} and analyze why certain cases can have an
extremely low or high latency.

% \begin{table}[t]
%   \caption{The transformation rules used for construction of equivalent Cypher queries from open-source projects.}
%   \label{tab:rules}
% \begin{tabular}{|l|p{5cm}|}
%     \hline
%     \textbf{Rule description} & \textbf{Graph algebra equation}  \\
%     \hline
%     Reversal of Direction & $
%     \downarrow_{y}^{x}(\circled{}_{y}) \equiv\uparrow_{x}^{y}(\circled{}_{x})
%     $\\
%     \hline
%     Graph Pattern Splitting & $
%     \downarrow_{y}^{z}(\downarrow_{y}^{x}(\circled{}_{y})
%     \equiv\downarrow_{y}^{x}(\uparrow_{z}^{y}(\circled{}_{z})))
%     $\\
%     \hline
%     Variable Replacement  & $\circled{}_{x} \equiv \circled{}_{y}$ \\
%     \hline
% \end{tabular}
% \end{table}
\iffalse
\begin{table}[t]
  \caption{The transformation rules used for construction of equivalent Cypher queries from open-source projects.}
  \label{tab:rules}
\begin{tabular}{|l|p{5cm}|}
    \hline
    \textbf{Rule description} & \textbf{Cypher example}  \\
    \hline
    Reversal of Direction & 
    \texttt{(n1)-[]->(n2)} $\equiv$  \texttt{(n2)-[]->(n1)}\\
    \hline
    Graph Pattern Splitting & 
    \texttt{(n1)-[]->(n2)<-[]-(n3)}
    $\equiv$\texttt{(n1)-[]->(n2), (n3)-[]->(n2)} \\
    \hline
    Variable Replacement  & \texttt{(x:L)} $\equiv$ \texttt{(x1:L)} \\
    \hline
\end{tabular}
\end{table}
\fi

\subsection{Dataset Construction}
\label{sec:evaluation-dataset}

\review{\#2: There are some minor inconsistencies regarding the number of queries in the CyEqSet set between the submitted manuscript and what is stated on the github repository. However, when I reviewed the raw data the number of queries does match with the paper, so I guess it is a matter of updating the repository.}

\review{\#2: It might be nice to also add negative examples into the testing set. I understand that the soundness result states that these would have no effect on the proposed method, but this might serve as a good sanity check. }

Currently, there are no open-source datasets of equivalent Cypher queries
available for evaluating our approach. Therefore, we construct a dataset,
\dataset{}, of equivalent Cypher queries to evaluate \approachname{}.
Specifically, we construct \dataset{} through two approaches: (1) translating
the open-source equivalent SQL query pairs from Calcite \cite{2021calcite} into
equivalent Cypher query pairs and (2) constructing equivalent Cypher query pairs
by rewriting Cypher queries using existing equivalent rewriting rules.

\textbf{Translation of SQL Calcite dataset.}
Calcite dataset \cite{2021calcite} that contains 232 pairs of equivalent SQL
queries is widely applied for evaluating SQL query equivalence
\cite{chu2017cosette,chu2018udp,zhou2019equitas,zhou2020spes,wang2022wetune,ding2023sqlsolver}.
Researchers have proposed SQL-to-Cypher translating tools, e.g., the
openCypherTranspiler \cite{openCypherTranspiler}. However, these tools only
support limited Cypher features, resulting in poor effectiveness when
translating the Calcite dataset. Therefore, we manually translate each SQL query
pair in Calcite dataset into a Cypher query pair and check the equivalence of
the obtained Cypher query pairs based on the principles of
Cytosm \cite{steer2017cytosm}.

However, since the syntax of Cypher is significantly different from SQL, the
following cases cannot be translated from SQL queries to Cypher queries. (1) We
discard SQL queries having sorting operations inside and outside subclauses,
since Cypher does not guarantee that the sorting of subqueries will be
preserved. (2) We discard SQL queries adopting operations on the results of
UNION or UNION ALL, which is not supported by Cypher. (3) We discard SQL queries
containing \texttt{GROUP BY} that cannot be represented as \texttt{DISTINCT} in
Cypher. Finally, we obtain 77 equivalent Cypher query pairs from Calcite in
total.

\iffalse
\begin{table}[t]
  \centering
  \caption{Equivalent Cypher queries transformed by 
  rewriting rules.}
  \vspace{-5pt}
  \label{tab:rules_result}
  \begin{tabular}{|l|l|l|l|l|}
    \hline
    \textbf{Rule} & \textbf{LDBC} & \makecell[l]{\textbf{Cypher-}\\\textbf{for-gremlin}} & \makecell[l]{\textbf{Graphdb-}\\\textbf{benchmarks}} & \textbf{Total}\\
    \hline
    RV & 7 & 14 & 15 & 36\\
    \hline
    RP & 4 & 5 & 12 & 21\\
    \hline
    SP & 2 & 4 & 5 & 11\\
    \hline
    Total & 13 & 23 & 32 & 68\\
    \hline
  \end{tabular}
\end{table}
\fi

\iffalse
Then, we apply the existing three equivalent Cypher rewriting rules on these Cypher queries as follows. (1) Renaming
variables (RV for short): This rule creates syntactically different Cypher queries by renaming the variable names of node
and relationship patterns. (2) Reversing path direction (RP
for short): This rule reverses the relationship patterns without
changing their incoming and outgoing node patterns. If the
Cypher query defines a graph pattern that contains a path with
multiple relationship patterns, we reverse the entire path. (3)
Splitting graph pattern (SP for short): This rule splits multiple relationship patterns in a Cypher graph pattern. For example, a graph pattern (n1)-[r1]->(n2)<-[r2]-(n4)
is splitted as (n1)-[r1]->(n2), (n2)<-[r2]-(n4)
\fi

\textbf{Equivalent query transformation through rewriting rules.} Since SQL
queries do not have some specific Cypher features, e.g., arbitrary-length paths,
the collected dataset via translating SQL queries cannot cover these Cypher
features. Therefore, to enrich and expand our dataset, we first collect
real-world Cypher queries from open-source projects, i.e.,
LDBC-snb-interactive-v1-impls \cite{erling2015ldbc} (a GDB benchmark),
Cypher-for-gremlin \cite{opencypher_cypher_for_gremlin} (a Cypher query plugin),
and Graphdb-benchmarks \cite{marco2018gdbbenchmark} (a GDB benchmark), and we
obtain 36 Cypher queries. Then, we apply three exiting Cypher
rewriting rules on these Cypher queries as follows. (1) \textbf{Renaming
variables}, which renames the variable names of node and relationship patterns.
(2) \textbf{Reversing path direction}, which reverses the relationship patterns
without changing their incoming and outgoing node patterns. (3)
\textbf{Splitting graph pattern}, which splits multiple relationship patterns in
a Cypher graph pattern.

We construct equivalent Cypher query pairs by applying these rules on each
real-world Cypher queries. Note that some Cypher queries can be successfully
rewrited by more than one rewriting rule, resulting in multiple equivalent
Cypher query pairs. Finally, we generated 68 equivalent Cypher query pairs.

In total, we construct \dataset{} with 148 equivalent Cypher query pairs from
the above two approaches. \dataset{} is representative since it covers various
of Cypher query fragments: \texttt{MATCH}, \texttt{OPTIONAL MATCH},
\texttt{RETURN}, \texttt{WHERE}, \texttt{WITH}, \texttt{UNION}, \texttt{UNION
ALL}, \texttt{UNWIND}, \texttt{LIMIT}, \texttt{SKIP}, \texttt{ORDER BY},
\texttt{ASC}, \texttt{DESC}, and aggregate functions (\texttt{SUM},
\texttt{COUNT}, \texttt{AVG}, \texttt{DISTINCT}, \texttt{MIN}, \texttt{MAX},
\texttt{COLLECT}). \dataset{} also covers advanced or complex Cypher
features, including all the features in \tabref{tab:all-summary} along with
special value (e.g., null, false) and queries that always return empty results.

\revision{To test the effectiveness of \approachname{} on proving non-equivalent
Cypher queries, we construct a dataset \texttt{CyNeqSet} containing 148
non-equivalent Cypher query pairs. Specifically, we mutate the equivalent Cypher
query pairs in \dataset{} by randomly applying one of the following mutation
rules: 1) changing the direction of a path, 2) changing the property values or
labels of some variables, 3) changing \texttt{UNION ALL} to \texttt{UNION} or
vice versa, 4) changing the value of \texttt{LIMIT} or \texttt{ORDER BY} and 5)
removing or adding \texttt{DISTINCT}. We further manually confirmed that each
pair of Cypher queries in \dataset{} is not equivalent.}\id{Q18}

\renewcommand{\arraystretch}{1}
\begin{table}[t]
  \centering
  \caption{Proved query pairs by \approachname{}.}
  \vspace{-5pt}
  \label{tab:verify}
  \begin{tabular}{|l|l|l|}
    \hline
    \textbf{Project} &  \makecell[l]{\textbf{Query pairs}} &
    \makecell[l]{ \textbf{Proved}}\\
    \hline
    Calcite-Cypher & 80 & 73\\
    \hline
    LDBC & 13 & 13\\
    \hline
    Cypher-for-gremlin & 23 & 23\\
    \hline
    Graphdb-benchmarks & 32 & 29\\
    \hline
    Total & 148 & 138\\
    \hline
  \end{tabular}
  \vspace{-10pt}
\end{table}

\subsection{Verification Result}
\label{sec:evaluation-capability}

\iffalse
\review{Reviewer C: Just showing the numbers in figure 4, without explaining
what is happening where and without splitting the total number in the components
that affect the overall performance we cannot actually understand what is going
on.\\
Overall, I feel that the evaluation is rather weak, without showing the impact
of each individual component, and a detailed analysis.}
\fi

\tabref{tab:verify} shows the evaluation result of \approachname{} on
\texttt{CyEqSet}. Specifically, out of 148 pairs of equivalent Cypher queries in
\texttt{CyEqSet}, \approachname{} can successfully prove the
equivalence of 138 query pairs. On the dataset translated from Calcite,
we proved 73 out of 80 (89\%) equivalent Cypher query pairs. On the dataset
constructed from real-world projects, we proved 65 out of 68 (approximately
96\%) equivalent Cypher query pairs. We further analyze the failed cases and
identify the reasons for these failed cases.

\iffalse
We evaluated the \approachname{}'s verification capabilities on Cypher query
pairs that have advanced features. Our result shows that \approachname{}
successfully proves all these Cypher query pairs.
\fi
\iffalse
\textbf{Characterizing results}. We characterize all equivalent Cypher query
pairs by the complex and advanced features and show the verification capability of
\approachname{}on these features as follows.
\begin{enumerate} [leftmargin=15pt]
  \item \textbf{Optional graph patterns}. Optional graph patterns are defined by
  \texttt{OPTIONAL MATCH} in Cypher. \approachname{} has proven all the 3
  equivalent Cypher query pairs with optional graph patterns.
  \item \textbf{Sorting operations}. Cypher query results are sorted by \texttt{ORDER BY}
  and \texttt{LIMIT}. \approachname{} has proven all the 14 equivalent Cypher
  query pairs with sorting operations and limit operations
  \item \textbf{Undirected relationships}. Undirected relationships are defined
  as \texttt{-[]-} in Cypher. \approachname{} has proven all the 8 equivalent
  Cypher query pairs with undirected relationship definitions.
  \item \textbf{Variable-length relationships}. Variable-length relationships
  are defined as \texttt{-[*1..2]-} in Cypher. \approachname{} has proven all
  the 10 equivalent Cypher query pairs with variable-length relationship definitions.
\end{enumerate}
\fi

\begin{itemize}[leftmargin = 12pt]
  \item \textbf{Sorting and truncation.} \approachname{} cannot prove equivalent 
  queries that contain inconsistent numbers of \texttt{ORDER
  BY...LIMIT...SKIP...} fragments within subqueries due to the limitation of our
  divide-and-conquer proving approach in \secref{sec:model-cypher-query}. 2
  cases failed due to this reason.
  \item \textbf{Aggregate.} \approachname{} cannot eliminate the intermediate
  variables in nested aggregates and aggregate computations, e.g.,
  \texttt{COUNT(SUM(n))} and \texttt{SUM(n)/COUNT(n)}. 4 cases failed due to
  this reason.
  \item \textbf{Uninterpreted function.} \approachname{} utilizes uninterpreted
  functions to model \texttt{COLLECT} aggregate and built-in functions,
  which cannot express their concrete semantics. 4 cases failed due to this
  reason.
\end{itemize}

\finding{finding1}{\approachname{} has successfully proved 138 out of 148 pairs
of equivalent Cypher queries, demonstrating its effectiveness.}

\revision{We evaluate \approachname{} on \texttt{CyNeqSet}, and check
whether \approachname{} can prove the non-equivalence of these pairs in
\texttt{CyNeqSet}. The experimental results show that \approachname{} proves all
test cases in \texttt{CyNeqSet} to be non-equivalent.}\id{Q18}

\iffalse
To illustrate the how tow Cypher queries are verified equivalent by
\approachname{} through our \uexpg{} construction, we provide a simplified
example from \dataset{}:
\begin{lstlisting}[ floatplacement=H,style=cypherstyle, caption={An illustrative Cypher query pair for constructing equivalent \uexpg{}.},mathescape=true, label=list:faled1]
/* Cypher query Q1 */
MATCH (n1)-[r1]->(n2)-[r2]->(n3) return n

/* Cypher query Q2 */
MATCH (n1)-[r1]->(n2), (n2)-[r2]->(n3) return n
\end{lstlisting}
In this test case, we first construct \uexpg{} for $Q1$ and $Q2$, that
\begin{align*}
  f_1(t)=&\sum_{e_1,e_2,e_3,e_4,e_5}[t=e_1]\times Node(e_1)\times Rel(e_2)\times Node(e_3)\\
         & \times [out(e_2)=e_3] \times [in(e_2)=e_1]\\
         & \times Rel(e_4) \times Node(e_5)\times [in(e_4)=e_3]\times [out(e_4)=e_5]
\end{align*}
\begin{align*}
  f_2(t)=&\sum_{e_1,e_2,e_3,e_4,e_5}[t=e_1]\times Node(e_1)\times Rel(e_2)\times Node(e_3)\\
         & \times [out(e_2)=e_3] \times [in(e_2)=e_1] \times Node(e_3)\\
         & \times Rel(e_4) \times Node(e_5)\times [in(e_4)=e_3]\times [out(e_4)=e_5]
\end{align*}

The difference between $f_1(t)$ and $f_2(t)$ is that $f_2(t)$ double multiplies
$Node(e3)$ when constructing \uexpg{} for the two parts of the
Cypher graph pattern in $Q2$ separated by \texttt{","}.
Since our algebraic function $Node(e)$ is boolean, we eliminate the duplicated
multiplication, making $f_1(t)$ and $f_2(t)$ equivalent.
\fi
% \finding{finding2}{Our rule-based normalization can improve the performance of
% \approachname. \wensheng{Delete this?}}
\subsection{Performance}
\label{sec:evaluation-performance}

\begin{figure}[t]
  \centering
  \includegraphics[width=1.0\linewidth]{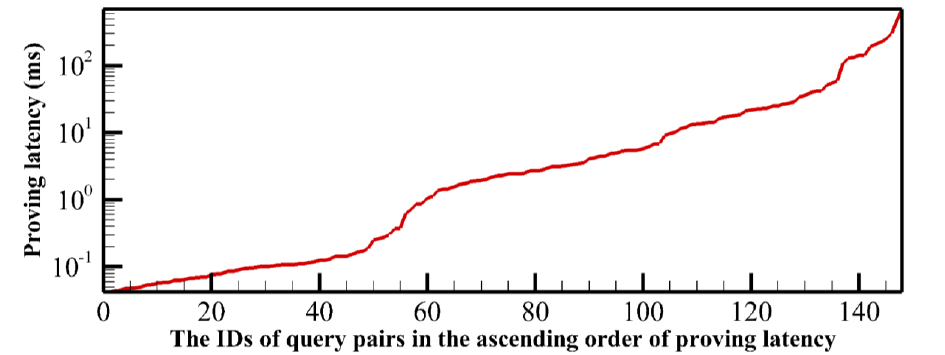}
  \vspace{-20pt}
  \caption{The proving latency of \approachname{}.}
  \vspace{-4mm}
  \label{fig:percentile}
\end{figure}

To evaluate the performance of \approachname{}, we conduct
tests about \approachname{}'s proving latency on a platform
equipped with an Intel Core i5-11300 processor and 16GB of RAM. 

The average latency for proving Cypher query equivalence across all test cases
in \texttt{CyEqSet} is 38ms. \figref{fig:percentile} shows the distribution of
proving latency for \approachname{}. 90\% of the test cases are proved by
\approachname{} in less than 100ms. Notably, 75 out of 148 Cypher query pairs
are verified with an exceptionally low latency of 10ms. Only 2 Cypher query
pairs require over 500ms latency for proving.

The Cypher query pair with the highest proving latency is due to the
\texttt{ORDER BY...LIMIT...} fragments within subqueries. We use the
divide-and-conquer based approach that proves their equivalence by dividing the
Cypher queries into subqueries by \texttt{ORDER BY...LIMIT...} fragments. Each
subquery is then individually proved for equivalence, leading to significantly
higher proving latency than other cases. 

\iffalse
\textbf{The returning mapping for Cypher queries.} Result element mapping
in \secref{sec:prove-uexpr-equiv} can affect the proving performance. We
investigate the mapping cases of returning elements of two Cypher queries before
sending them to the SMT sovler. There are 83 pairs of Cypher queries that need
result element mapping. Out of these, 49 cases produce a single mapping and are
proven successfully. 29 cases generate fewer than 20 mappings, with the first
one being sufficient for a successful proof. Other cases with more mappings are
still proven successfully with the first one. 4 failed cases produce fewer than
7 mappings and are processed in less than 142.4 ms, as the translated queries
return similar, small sets of elements. However, our method cannot effectively
handle queries with a large number of unordered elements that are difficult to
distinguish. Additionally, 4 cases return inconsistent elements and always
produce empty results.
\fi

\iffalse
\review{Reviewer A: How do we know that the measured proving latency is acceptable? Are SQL query equivalence proofers in the same ballpark? Or is almost 90\% under 100ms just a nice sounding result? Or is it because humans get nervous after about 200ms when they have clicked a button and nothing happens? What is the benchmark here?}
\fi

\finding{finding4}{\approachname{} can efficiently prove the equivalence of
Cypher queries with a latency of 38ms on average.}

\section{Discussion}
In this section, we first discuss the limitations of \approachname{}.
Then, we discuss how to extend \approachname{} for supporting other graph
query languages, \revision{e.g., GQL \cite{GQLISO, GQLgraphPattern}, Gremlin
\cite{gremlin} and SPARQL \cite{perez2009sparql}}\id{Q11}. 

\review{\#3: D2. What are the potential challenges in extending GraphQE to support advanced Cypher features like CALL or user-defined functions?}

\subsection{Limitations}
\revision{Although \approachname{} is effective in proving the equivalence of Cypher
queries, \approachname{} is unable to adequately support some Cypher features,
which necessitate the development of new approaches in the future.}
\begin{itemize}[leftmargin = 12pt]
  \item \revision{\approachname{} cannot prove the equivalence of Cypher queries
  containing nested aggregates and aggregation computations, e.g.,
  \texttt{SUM(SUM(n))} and \texttt{SUM(n)/COUNT(n)}, because their
  corresponding \uexpg{}s contain intermediate variables that cannot be eliminated
  through our approach.}
  \item \revision{\approachname{} cannot prove the equivalence of Cypher
  queries containing different number of \texttt{ORDER BY...LIMIT...SKIP...} fragments,
  because our divide-and-conquer proving process requires that two equivalent
  Cypher queries have the same number of \texttt{ORDER BY...LIMIT...SKIP...} fragments.}
  \item \revision{\approachname{} cannot model the concrete semantics of
  built-in functions and user-defined functions, and cannot support some Cypher
  features, e.g., \texttt{YIELD} in \texttt{CALL}.}\id{Q19, Q23}
\end{itemize}

\subsection{Extending \approachname{} to other Graph Query Languages}
\review{\#3: D4. The discussion on how GraphQE could be adapted to support GQL, as well as other graph query languages such as Gremlin and SPARQL would benefit from being further detailed.}
\review{\#1: O4. Why is GQL not considered? There is a substantial overlap between Cypher and GQL. As far as I can see, the main difference is the edge-injective semantics in Cypher, which is not required in GQL.}

\label{sec:discussion-generalization}
\iffalse
Unlike relational database systems, graph database systems have long lacked a consensus on a
graph query language. Graph query languages widely used by GDBs include Cypher,
Gremlin, and SPARQL, which is designed for RDF graph storage systems, among
others. In April of this year, ISO released Graph Query Language (GQL)
\cite{GQLISO}, the first standardized general graph query language designed
based on and compatible with Cypher. GQL follows the same graph query paradigm
as Cypher and adopts Cypher-like syntax while extending Cypher's functionality
with additional features. However, few graph databases currently support it. We
will now discuss how \approachname{} provides support for the aforementioned
graph query languages.
\fi

Although \approachname{} is designed for Cypher queries, it can be extended to
other graph query languages, \revision{e.g., GQL \cite{GQLISO, GQLgraphPattern},
Gremlin \cite{gremlin} and SPARQL \cite{perez2009sparql}}\id{Q11}. Next, we
introduce how to extend \approachname{} to support these graph query
languages.

\revision{\textbf{Supporting GQL.} GQL is a new standard graph query language
that extends Cypher and is fully compatible with Cypher. We can easily extend
\approachname{} to support GQL from the following aspects. First,
\approachname{} needs to support the property graph model adopted by GQL.
\approachname{} needs to remove its relationship-injective semantics when
constructing \uexpg{}s for GQL. \approachname{} needs to allow a relationship to
be specified with more than one label. Second, \approachname{} can support
undirected relationships using uninterpreted function. 
\iffalse
For example, for an undirected relationship $r1$ defined by
\texttt{(n1)-[r1]-(n2)}, \approachname{} models it as $Node(n_1)\times
Rel(r_1)$$\times Node(n_2)$$\times UNDIRECTED(r_1)\times([in(r_1)=$
$n_1]$$\times [out(r_1)=Node(n_2)]+[in(r_1)=n_2]\times [out(r_1)=Node(n_1)])$.
\fi
Third, \approachname{} needs to model the new introduced features in GQL, e.g.,
\texttt{EXCEPT}, \texttt{FILTER} and \texttt{FOR}.}\id{Q2, Q8, Q19, Q25}

\iffalse
\textbf{Generalizing \approachname{} to Gremlin.} Gremlin is a functional query
language based on property graph databases. The major difference among Gremlin
and Cypher is their syntaxes. The major difference among Gremlin and Cypher is
their syntaxes. Gremlin adopts graph pattern matching based on homomorphism
without the restrictions of no-repeated-edge semantics, we can still easily
model them by simply removing no-repeated-edge semantic expressions in
G-expressions. 
\fi

\revision{\textbf{Supporting Gremlin.} Gremlin is designed as a functional query
language on property graphs. For example, the Gremlin query
$g.V().hasLabel(``person").has(``name", ``Alice")$
$.out(``knows").values(``name")$ finds all people that Alice knows. The semantic
difference between Gremlin and Cypher is minimal. Gremlin does not use
relationship-injective semantics and allows a relationship to have more than one
label. \approachname{} needs to remove its relationship-injective semantics
while constructing \uexpg{}s and allows a relationship to be specified with more
than one label. Then, \approachname{} needs to model the Gremlin fragments as
\uexpg{}s.}\id{Q25}

\iffalse
\textbf{Generalizing \approachname{} to SQARQL.} Generalizing GraphQE to the RDF
graph model and SPARQL \cite{perez2009sparql} may face more challenges, as the
RDF graph model offers greater flexibility, and the extended functionality
introduced by RDF* \cite{RDFstar} increases the complexity of modeling. For example
relationships between subgraphs cannot be modeled by GraphQE. However, we can
still directly model the basic cases of RDF graph model by treat RDF graphs as
special property graphs \cite{angles2017foundations}. In this way, predicate
URLs are modeled as relationships, while their corresponding subjects and
objects are treated as nodes. 
\fi

\revision{\textbf{Supporting SPARQL.} SPARQL is a graph query
language designed for RDF graphs \cite{wylot2018rdf}. SPARQL defines graph
patterns as triples $\langle Subject, Predicate, Object \rangle$ that match
paths from $Subject$ to $Object$ through $Predicate$. We can model SPARQL
queries on an RDF graph in the same way of modeling Cypher queries, since an RDF
graph can be treated as a special kind of property graphs
\cite{angles2017foundations}. \approachname{} needs to fix the gap between the RDF graph
model and property graph model, and build \uexpg{}s for SPARQL
fragments.}\id{Q25}

\review{\#3: O1. The applicability of the proposed framework is restricted. The paper would benefit from a more in depth discussion on the technical limitations of the method, e.g., support for nested aggregates, and how these can be addressed, as well as on the needed extensions to support GQL.}
\vspace{-0.5mm} 

\section{Related Work}
\vspace{-0.5mm}
\textbf{SQL query equivalence provers}. Researchers have proposed syntax-based
\cite{chu2017cosette,chu2018udp,zhou2019equitas,zhou2020spes} and
semantics-based \cite{wang2022wetune,ding2023sqlsolver,wang2024QED} SQL
equivalence provers. Syntax-based approaches prove the equivalence of SQL
queries by checking the isomorphism of SQL algebraic representations. For
example, SPES \cite{zhou2020spes} proposes tree-based SQL algebraic
representations to model SQL relaional algebras, and adopts Z3 SMT solver to
prove their equivalence. However, these approaches cannot handle equivalent SQL
queries that have significantly different syntactic structures. Semantics-based
approaches model SQL queries using semiring expressions, and transform them into
first-order logic expressions that can be verified by SMT solvers. However,
these approaches cannot be directly used to prove Cypher query equivalence.

\textbf{Equivalent graph queries and their applications}. Equivalent rewriting
of graph queries is essential in both bug detection of GDBs
\cite{Kamm2023Gdbmeter,jiang2023graphgenie,zhuang2023testing} and query
optimization \cite{holsch2016graphpattern, choudhury2014query}. Kamm et al.
\cite{Kamm2023Gdbmeter} test Gremlin-supported GDBs via constructing equivalent
Gremlin queries as test oracles through equivalent query rewriting. Jiang et al.
\cite{jiang2023graphgenie} use equivalent rewriting to test Cypher-supported
GDBs. However, these rewriting rules are manually defined. Our work aims to
propose an automated graph query equivalence prover.
\vspace{-0.5mm}
\section{Conclusion}
\vspace{-0.5mm}
Query equivalence proving is a fundamental problem in database research.
However, we still lack a graph query equivalence prover for the emerging graph
databases. In this paper, we propose \approachname{}, the first Cypher query
equivalence prover, to determine whether two Cypher queries are semantically
equivalent. We model Cypher queries as \uexpg{}s, and then prove their
equivalence by using constraint solvers. We further construct a dataset that
contains 148 pairs of equivalent Cypher queries. We evaluate \approachname{} on
this dataset, and \approachname{} has proved 138 pairs of equivalent Cypher
queries, demonstrating the effectiveness of \approachname{}.

\section*{Acknowledgments}

This work was partially supported by National Natural Science Foundation of
China (62072444), Major Project of ISCAS (ISCAS-ZD-202302), Basic Research
Project of ISCAS (ISCAS-JCZD-202403), and Youth Innovation Promotion Association
at Chinese Academy of Sciences (Y2022044).

\balance
\bibliographystyle{IEEEtranS}
\bibliography{CypherEquiv}

% Generated by IEEEtranS.bst, version: 1.12 (2007/01/11)
\begin{thebibliography}{10}
\providecommand{\url}[1]{#1}
\csname url@samestyle\endcsname
\providecommand{\newblock}{\relax}
\providecommand{\bibinfo}[2]{#2}
\providecommand{\BIBentrySTDinterwordspacing}{\spaceskip=0pt\relax}
\providecommand{\BIBentryALTinterwordstretchfactor}{4}
\providecommand{\BIBentryALTinterwordspacing}{\spaceskip=\fontdimen2\font plus
\BIBentryALTinterwordstretchfactor\fontdimen3\font minus
  \fontdimen4\font\relax}
\providecommand{\BIBforeignlanguage}[2]{{%
\expandafter\ifx\csname l@#1\endcsname\relax
\typeout{** WARNING: IEEEtranS.bst: No hyphenation pattern has been}%
\typeout{** loaded for the language `#1'. Using the pattern for}%
\typeout{** the default language instead.}%
\else
\language=\csname l@#1\endcsname
\fi
#2}}
\providecommand{\BIBdecl}{\relax}
\BIBdecl

\bibitem{azurecosmos}
\BIBentryALTinterwordspacing
``Azure cosmos db.'' [Online]. Available:
  \url{https://azure.microsoft.com/en-us/services/cosmos-db/}
\BIBentrySTDinterwordspacing

\bibitem{opencypher}
\BIBentryALTinterwordspacing
``Opencypher documentation.'' [Online]. Available:
  \url{https://opencypher.org/}
\BIBentrySTDinterwordspacing

\bibitem{janusgraph2019}
\BIBentryALTinterwordspacing
``Janusgraph: Official documentation,'' 2019. [Online]. Available:
  \url{https://docs.janusgraph.org}
\BIBentrySTDinterwordspacing

\bibitem{2021calcite}
\BIBentryALTinterwordspacing
``Calcite test suite,'' 2021. [Online]. Available:
  \url{"https://github.com/georgia-tech-db/Qizhou2020spes/blob/main/testData"}
\BIBentrySTDinterwordspacing

\bibitem{arangodb}
``{ArangoDB}: {The} native multi-model database,''
  \url{"https://www.arangodb.com/"}, 2023.

\bibitem{dbengines2023graph}
``Graph database ranking,''
  \url{"https://db-engines.com/en/ranking/graph+dbms"}, 2023.

\bibitem{gremlin}
``Gremlin: {The} graph traversal machine and language,''
  \url{https://tinkerpop.apache.org/gremlin.html}, 2023.

\bibitem{opencypher_cypher_for_gremlin}
``Cypher for {Gremlin},''
  \url{https://github.com/opencypher/cypher-for-gremlin}, 2024.

\bibitem{memgraph2023}
\BIBentryALTinterwordspacing
``Memgraph documentation,'' 2024. [Online]. Available:
  \url{https://memgraph.com/docs/memgraph/introduction}
\BIBentrySTDinterwordspacing

\bibitem{openCypherTranspiler}
``{openCypherTranspiler}: {An} open source project for transpiling cypher
  queries,'' \url{https://github.com/microsoft/openCypherTranspiler}, 2024.

\bibitem{angles2017foundations}
R.~Angles, M.~Arenas, P.~Barcel{\'o}, A.~Hogan, J.~Reutter, and D.~Vrgo{\v{c}},
  ``Foundations of modern query languages for graph databases,'' \emph{ACM
  Computing Surveys (CSUR)}, vol.~50, no.~5, pp. 1--40, 2017.

\bibitem{barcel2013querying}
P.~Barcel{\'o}~Baeza, ``Querying graph databases,'' in \emph{Proceedings of ACM
  SIGMOD-SIGACT-SIGAI Symposium On Principles of Database Systems (PODS)},
  2013, pp. 175--188.

\bibitem{marco2018gdbbenchmark}
M.~Brandizi, A.~Singh, and K.~Hassani-Pak, ``Getting the best of linked data
  and property graphs: {Rdf2neo} and the {KnetMiner} use case,'' in
  \emph{Proceedings of International Conference Semantic Web Applications and
  Tools for Life Sciences (SWAT4LS)}, 2018.

\bibitem{chandra1977optimal}
A.~K. Chandra and P.~M. Merlin, ``Optimal implementation of conjunctive queries
  in relational databases,'' in \emph{Proceedings of the ACM Symposium on
  Theory of Computing (STOC)}, 1977, pp. 77--90.

\bibitem{choudhury2014query}
S.~Choudhury, L.~Holder, G.~Chin, P.~Mackey, K.~Agarwal, and J.~Feo, ``Query
  optimization for dynamic graphs,'' \emph{arXiv preprint arXiv:1407.3745},
  2014.

\bibitem{chu2018udp}
S.~Chu, B.~Murphy, J.~Roesch, A.~Cheung, and D.~Suciu, ``Axiomatic foundations
  and algorithms for deciding semantic equivalences of {SQL} queries,''
  \emph{Proceedings of the VLDB Endowment (PVLDB)}, vol.~11, no.~11, pp.
  1482--1495, 2018.

\bibitem{chu2017cosette}
S.~Chu, K.~Weitz, A.~Cheung, and D.~Suciu, ``Hottsql: Proving query rewrites
  with univalent sql semantics,'' vol.~52, no.~6, 2017, pp. 510--524.

\bibitem{cohen1999rewriting}
S.~Cohen, W.~Nutt, and A.~Serebrenik, ``Rewriting aggregate queries using
  views,'' in \emph{Proceedings of the ACM Symposium on Principles of Database
  Systems (SIGMOD-SIGACT-SIGART)}, 1999, pp. 155--166.

\bibitem{demoura2008z3}
L.~De~Moura and N.~Bj{\o}rner, ``{Z3}: {An} efficient {SMT} solver,'' in
  \emph{Proceedings of International Conference on Tools and Algorithms for the
  Construction and Analysis of Systems (TACAS)}, 2008, pp. 337--340.

\bibitem{de2017smart}
R.~De~Virgilio, ``Smart rdf data storage in graph databases,'' in
  \emph{IEEE/ACM International Symposium on Cluster, Cloud and Grid Computing
  (CCGRID)}.\hskip 1em plus 0.5em minus 0.4em\relax IEEE, 2017, pp. 872--881.

\bibitem{GQLgraphPattern}
A.~Deutsch, N.~Francis, A.~Green, K.~Hare, B.~Li, L.~Libkin, T.~Lindaaker,
  V.~Marsault, W.~Martens, J.~Michels, F.~Murlak, S.~Plantikow, P.~Selmer,
  O.~van Rest, H.~Voigt, D.~Vrgo\v{c}, M.~Wu, and F.~Zemke, ``Graph pattern
  matching in gql and {SQL/PGQ},'' in \emph{Proceedings of the ACM
  International Conference on Management of Data (SIGMOD)}, 2022, pp.
  2246--2258.

\bibitem{ding2023sqlsolver}
H.~Ding, Z.~Wang, Y.~Yang, D.~Zhang, Z.~Xu, H.~Chen, R.~Piskac, and J.~Li,
  ``Proving query equivalence using linear integer arithmetic,''
  \emph{Proceedings of the ACM on Management of Data (SIGMOD)}, vol.~1, no.~4,
  pp. 1--26, 2023.

\bibitem{eckman2006graph}
B.~A. Eckman and P.~G. Brown, ``Graph data management for molecular and cell
  biology,'' \emph{IBM Journal of Research and Development}, vol.~50, no.~6,
  pp. 545--560, 2006.

\bibitem{erling2015ldbc}
O.~Erling, A.~Averbuch, J.~Larriba-Pey, H.~Chafi, A.~Gubichev, A.~Prat, M.-D.
  Pham, and P.~Boncz, ``The {LDBC} social network benchmark: {Interactive}
  workload,'' in \emph{Proceedings of ACM International Conference on
  Management of Data (SIGMOD)}, 2015, pp. 619--630.

\bibitem{fan2019social}
W.~Fan, Y.~M. 0001, Q.~Li, Y.~He, Y.~E. Zhao, J.~Tang, and D.~Yin, ``Graph
  neural networks for social recommendation,'' in \emph{Proceedings of the
  World Wide Web Conference (WWW)}, 2019, pp. 417--426.

\bibitem{farber2012sap}
F.~F{\"a}rber, S.~K. Cha, J.~Primsch, C.~Bornh{\"o}vd, S.~Sigg, and W.~Lehner,
  ``Sap hana database: Data management for modern business applications,''
  \emph{ACM Sigmod Record}, vol.~40, no.~4, pp. 45--51, 2012.

\bibitem{francis2018Cypher}
N.~Francis, A.~Green, P.~Guagliardo, L.~Libkin, T.~Lindaaker, V.~Marsault,
  S.~Plantikow, M.~Rydberg, P.~Selmer, and A.~Taylor, ``Cypher: {An} evolving
  query language for property graphs,'' in \emph{Proceedings of ACM
  International Conference on Management of Data (SIGMOD)}, 2018, pp.
  1433--1445.

\bibitem{ganski1987optimization}
R.~A. Ganski and H.~K. Wong, ``Optimization of nsted sql queries revisited,''
  \emph{Proceedings of ACM International Conference on Management of Data
  (SIGMOD)}, vol.~16, no.~3, pp. 23--33, 1987.

\bibitem{green2007krelation}
T.~J. Green, G.~Karvounarakis, and V.~Tannen, ``Provenance semirings,'' in
  \emph{Proceedings of ACM SIGMOD-SIGACT-SIGART Symposium on Principles of
  Database Systems (PODS)}, 2007, pp. 31--40.

\bibitem{holsch2016graphpattern}
J.~H{\"{o}}lsch and M.~Grossniklaus, ``An algebra and equivalences to transform
  graph patterns in {Neo4j},'' in \emph{Proceedings of the Workshops of the
  EDBT/ICDT Joint Conference (EDBT/ICDT)}, 2016.

\bibitem{imarc2024graph}
{IMARC Group}, ``Graph database market size, share, growth report 2024-32,''
  \url{www.imarcgroup.com}, 2024.

\bibitem{GQLISO}
\emph{Information technology — Database languages — GQL}, International
  Organization for Standardization (ISO) Std. ISO/IEC FDIS 39\,075:2024, 2024.

\bibitem{jiang2023graphgenie}
Y.~Jiang, J.~Liu, J.~Ba, R.~H.~C. Yap, Z.~Liang, and M.~Rigger, ``Detecting
  logic bugs in graph database management systems via injective and surjective
  graph query transformation,'' in \emph{Proceedings of IEEE/ACM International
  Conference on Software Engineering (ICSE)}, 2023, pp. 531--542.

\bibitem{Kamm2023Gdbmeter}
M.~Kamm, M.~Rigger, C.~Zhang, and Z.~Su, ``Testing graph database engines via
  query partitioning,'' in \emph{Proceedings of ACM SIGSOFT International
  Symposium on Software Testing and Analysis (ISSTA)}, 2023, pp. 140--149.

\bibitem{kemper2015beginning}
C.~Kemper, \emph{Beginning Neo4j}.\hskip 1em plus 0.5em minus 0.4em\relax
  Springer, 2015.

\bibitem{kim2012multiplicative}
M.~Kim and J.~Leskovec, ``Multiplicative attribute graph model of real-world
  networks,'' \emph{Internet Mathematics}, vol.~8, no. 1-2, pp. 113--160, 2012.

\bibitem{levatich2020solving}
M.~Levatich, N.~Bj{\o}rner, R.~Piskac, and S.~Shoham, ``Solving using
  approximations,'' in \emph{International Conference on Verification, Model
  Checking, and Abstract Interpretation (VMCAI)}, 2020, pp. 360--378.

\bibitem{marton2017formalising}
J.~Marton, G.~Sz{\'a}rnyas, and D.~Varr{\'o}, ``Formalising opencypher graph
  queries in relational algebra,'' in \emph{Proceedings of the Advances in
  Databases and Information Systems (ADBIS)}, 2017, pp. 182--196.

\bibitem{mpinda2015evaluation}
S.~A.~T. Mpinda, L.~C. Ferreira, M.~X. Ribeiro, and M.~T.~P. Santos,
  ``Evaluation of graph databases performance through indexing techniques,''
  \emph{International Journal of Artificial Intelligence \& Applications
  (IJAIA)}, vol.~6, no.~5, pp. 87--98, 2015.

\bibitem{parr2013antlr4}
T.~Parr, ``The definitive {ANTLR4} reference,'' \emph{The Definitive ANTLR4
  Reference}, pp. 1--326, 2013.

\bibitem{perez2009sparql}
J.~P{\'e}rez, M.~Arenas, and C.~Gutierrez, ``Semantics and complexity of
  {SPARQL},'' \emph{ACM Transactions on Database Systems (TODS)}, vol.~34,
  no.~3, pp. 1--45, 2009.

\bibitem{ren21frauddetection}
Y.~Ren, H.~Zhu, J.~Zhang, P.~Dai, and L.~Bo, ``{EnsemFDet}: {An} ensemble
  approach to fraud detection based on bipartite graph,'' in \emph{Proceedings
  of International Conference on Data Engineering (ICDE)}, 2021, pp.
  2039--2044.

\bibitem{ritter2021orientdb}
D.~Ritter, L.~Dell'Aquila, A.~Lomakin, and E.~Tagliaferri, ``Orientdb: A nosql,
  open source mmdms.'' in \emph{Proceedings of the British International
  Conference on Databases (BICOD)}, 2021, pp. 10--19.

\bibitem{sagiv1980equivalences}
Y.~Sagiv and M.~Yannakakis, ``Equivalences among relational expressions with
  the union and difference operators,'' \emph{Journal of the ACM (JACM)},
  vol.~27, no.~4, pp. 633--655, 1980.

\bibitem{steer2017cytosm}
B.~A. Steer, A.~Alnaimi, M.~A. Lotz, F.~Cuadrado, L.~M. Vaquero, and
  J.~Varvenne, ``{Cytosm}: {Declarative} property graph queries without data
  migration,'' in \emph{Proceedings of International Workshop on Graph
  Data-management Experiences and Systems (GRADES)}, 2017, pp. 1--6.

\bibitem{ullmann1976algorithm}
J.~R. Ullmann, ``An algorithm for subgraph isomorphism,'' \emph{Journal of the
  ACM (JACM)}, vol.~23, no.~1, pp. 31--42, 1976.

\bibitem{wang2020covid}
J.~Wang, K.~Wang, J.~Li, J.~Jiang, Y.~Wang, J.~Mei, and S.~Li, ``Accelerating
  epidemiological investigation analysis by using nlp and knowledge reasoning:
  A case study on covid-19,'' in \emph{American Medical Informatics Association
  Annual Symposium (AMIA)}, 2020.

\bibitem{wang2024QED}
S.~Wang, S.~Pan, and A.~Cheung, ``Qed: A powerful query equivalence decider for
  sql,'' in \emph{Proceedings of International Conference on Very Large Data
  Bases (PVLDB)}, 2024, pp. 3602--3614.

\bibitem{wang2022wetune}
Z.~Wang, Z.~Zhou, Y.~Yang, H.~Ding, G.~Hu, D.~Ding, C.~Tang, H.~Chen, and
  J.~Li, ``{WeTune}: {Automatic} discovery and verification of query rewrite
  rules,'' in \emph{Proceedings of ACM International Conference on Management
  of Data (SIGMOD)}, 2022, pp. 94--107.

\bibitem{wu2022nebula}
M.~Wu, X.~Yi, H.~Yu, Y.~Liu, and Y.~Wang, ``Nebula graph: {An} open source
  distributed graph database,'' \emph{CoRR}, vol. abs/2206.07278, 2022.

\bibitem{wylot2018rdf}
M.~Wylot, M.~Hauswirth, P.~Cudré-Mauroux, and S.~Sakr, ``Rdf data storage and
  query processing schemes: A survey,'' \emph{ACM Computing Surveys}, vol.~51,
  no.~4, 2018.

\bibitem{zheng2024qudi}
Y.~Zheng, W.~Dou, L.~Tang, Z.~Cui, Y.~Gao, J.~Song, L.~Xu, J.~Zhu, W.~Wang,
  J.~Wei, H.~Zhong, and T.~Huang, ``Testing gremlin-{Based} graph database
  systems via query disassembling,'' in \emph{Proceedings of ACM SIGSOFT
  International Symposium on Software Testing and Analysis (ISSTA)}, 2024, pp.
  1695--1707.

\bibitem{zheng2024dot}
Y.~Zheng, W.~Dou, L.~Tang, Z.~Cui, J.~Song, Z.~Cheng, W.~Wang, J.~Wei,
  H.~Zhong, and T.~Huang, ``Differential optimization testing of gremlin-based
  graph database systems,'' \emph{Proceedings of IEEE International Conference
  on Software Testing, Verification and Validation (ICST)}, pp. 25--36, 2024.

\bibitem{zhou2019equitas}
Q.~Zhou, J.~Arulraj, S.~Navathe, W.~Harris, and D.~Xu, ``Automated verification
  of query equivalence using satisfiability modulo theories,'' vol.~12, no.~11,
  2019, pp. 1276--1288.

\bibitem{zhou2020spes}
Q.~Zhou, J.~Arulraj, S.~B. Navathe, W.~Harris, and J.~Wu, ``Spes: {A} symbolic
  approach to proving query equivalence under bag semantics,'' in \emph{IEEE
  International Conference on Data Engineering (ICDE)}, 2022, pp. 2735--2748.

\bibitem{zhuang2023testing}
Z.~Zhuang, P.~Li, P.~Ma, W.~Meng, and S.~Wang, ``Testing graph database systems
  via graph-aware metamorphic relations,'' \emph{Proceedings of the VLDB
  Endowment (PVLDB)}, vol.~17, no.~4, pp. 836--848, 2023.

\end{thebibliography}

\end{document}